\documentclass[usenatbib,amsmath,amsfonts]{mn2e}

\usepackage[dvips]{graphicx}
\usepackage{amssymb}
\usepackage{txfonts}
\usepackage{colordvi}

\newcommand{\integral}{{\textit{INTEGRAL}}}
\newcommand{\xte}{{\textit{RXTE}}}

\newcommand{\sax}{{\textit{Beppo\-SAX}}}
\newcommand{\gro}{{\textit{CGRO}}}
\newcommand{\fermi}{{\textit{Fermi}}}
\newcommand{\agile}{{\textit{AGILE}}}
\newcommand{\swift}{{\textit{Swift}}}
\newcommand{\msun}{{\rm M}_{\sun}}
\newcommand{\rsun}{{\rm R}_{\sun}}

\newcommand{\g}{$\gamma$}

\newbox\grsign \setbox\grsign=\hbox{$>$} \newdimen\grdimen \grdimen=\ht\grsign
\newbox\simlessbox \newbox\simgreatbox \newbox\simpropbox
\setbox\simgreatbox=\hbox{\raise.5ex\hbox{$>$}\llap
     {\lower.5ex\hbox{$\sim$}}}\ht1=\grdimen\dp1=0pt
\setbox\simlessbox=\hbox{\raise.5ex\hbox{$<$}\llap
     {\lower.5ex\hbox{$\sim$}}}\ht2=\grdimen\dp2=0pt
\setbox\simpropbox=\hbox{\raise.5ex\hbox{$\propto$}\llap
     {\lower.5ex\hbox{$\sim$}}}\ht2=\grdimen\dp2=0pt
\def\ga{\mathrel{\copy\simgreatbox}}
\def\la{\mathrel{\copy\simlessbox}}

\title[Jet contributions to the spectrum of Cyg X-1]{Jet contributions to the broad-band spectrum\\ of Cyg X-1
in the hard state}

\author[A. A. Zdziarski et al.]
{Andrzej A. Zdziarski,$^1$ Patryk Pjanka$^2$, Marek Sikora$^1$ and {\L}ukasz Stawarz$^{3,4}$\\
$^1$Centrum Astronomiczne im.\ M. Kopernika, Bartycka 18, PL-00-716 Warszawa, Poland\\
$^2$Obserwatorium Astronomiczne Uniwersytetu Warszawskiego, Al. Ujazdowskie 4, 00-478 Warszawa, Poland\\
$^3$Institute of Space and Astronautical Science JAXA, 3-1-1 Yoshinodai, Chuo-ku, Sagamihara, Kanagawa 252-5210, Japan\\
$^4$Astronomical Observatory, Jagiellonian University, Orla 171, 30-244 Krak{\'o}w, Poland
}

\date{Accepted 2014 May 19. Received 2014 May 16; in original form 2014 March 19}

\pagerange{\pageref{firstpage}--\pageref{lastpage}}
\pubyear{2013}

\begin{document}

\maketitle

\label{firstpage}

\begin{abstract}
We apply the jet model developed in the preceding paper of Zdziarski et al.\ to the hard-state emission spectra of Cyg X-1. We augment the model for the analytical treatment of the particle evolution beyond the energy dissipation region, and allow for various forms of the acceleration rate. We calculate the resulting electron and emission spectra as functions of the jet height, along with the emission spectra integrated over the outflow. The model accounts well for the observed radio, infrared, and GeV fluxes of the source, although the available data do not provide unique constraints on the model free parameters. The contribution of the jet emission in the UV--to--X-ray range turns out to be in all the cases negligible compared to the radiative output of the accretion component. Nevertheless, we find out that it is possible to account for the observed flux of Cyg X-1 at MeV energies by synchrotron jet emission, in accord with the recent claims of the detection of strong linear polarization of the source in that range. However, this is possible only assuming a very efficient particle acceleration leading to the formation of flat electron spectra, and jet magnetic fields much above the equipartition level.
\end{abstract}

\section{Introduction}
\label{intro}

\begin{figure*}
\centerline{\includegraphics[width=11.5cm]{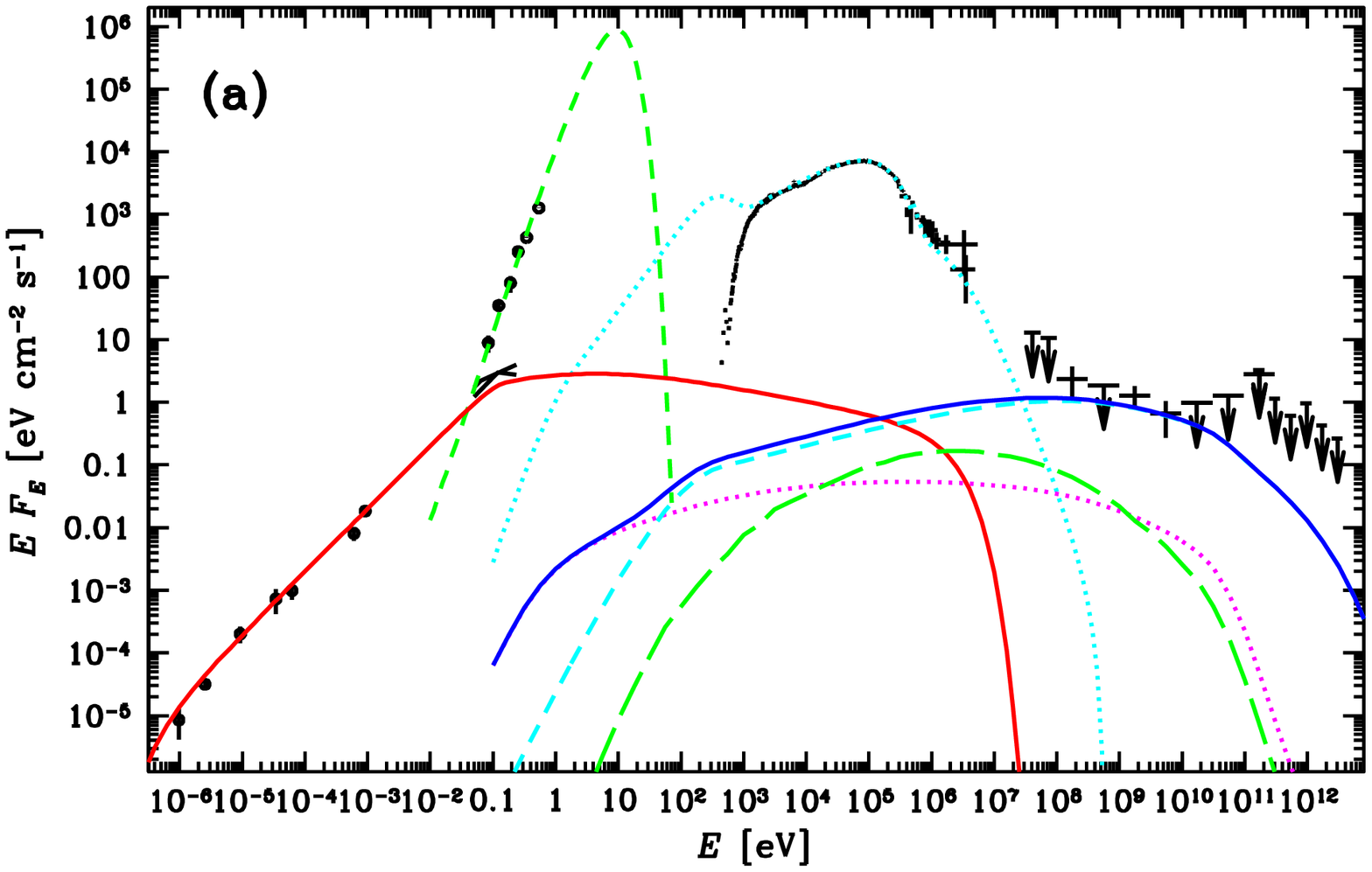}} 
\centerline{\includegraphics[width=11.5cm]{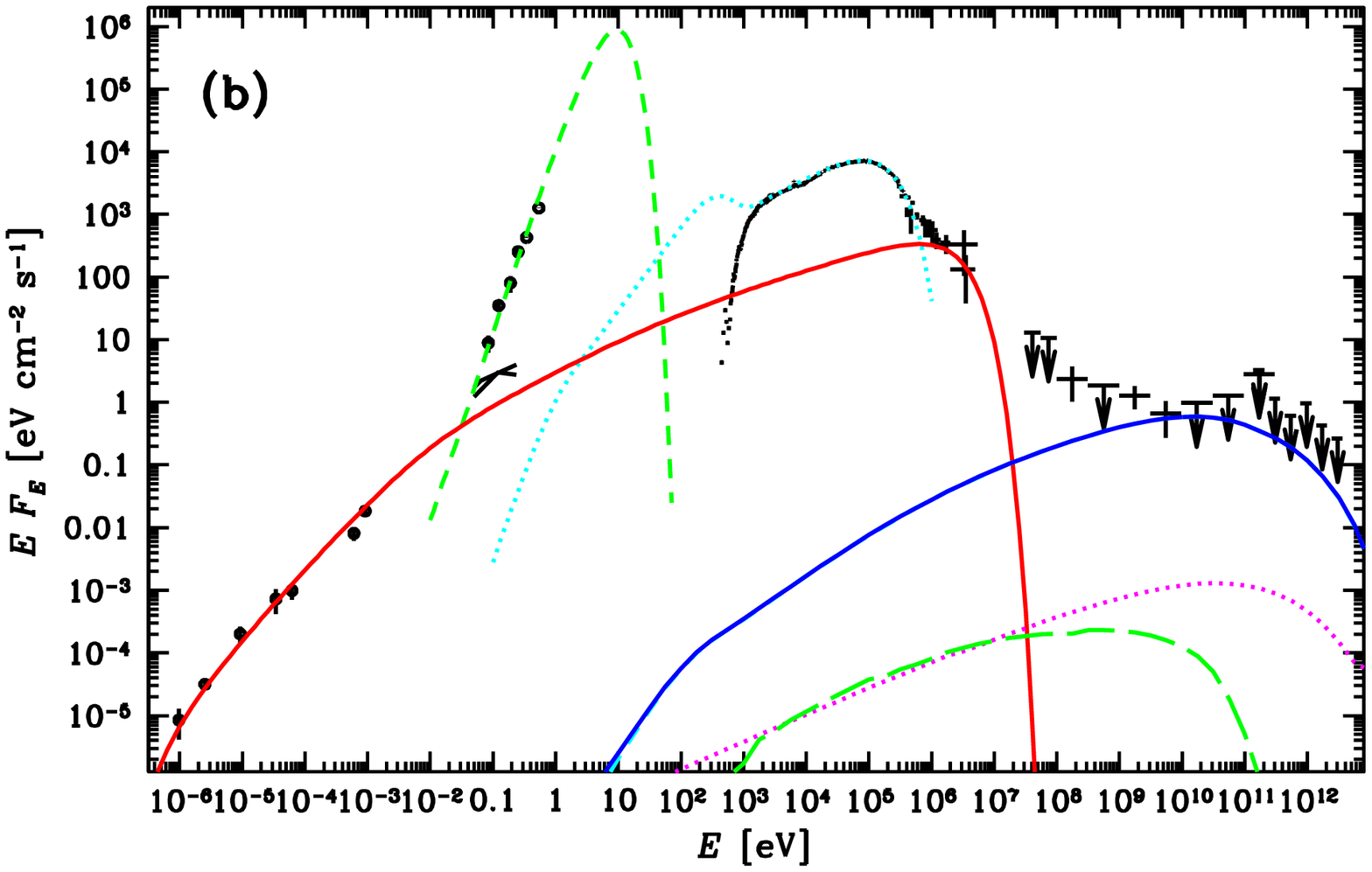}} 
\caption{The hard-state broad-band spectrum of Cyg X-1 shown together with our jet models (a) 1 and (b) 2. Fluxes in the radio/mm range from \citet{pandey07} and \citet{fender00}, along with the total IR fluxes \citep{persi80,mirabel96}, are denoted with black circles. The IR spectra of the jet component from \citet{rahoui11} are given as black curves. Small black squares correspond to the X-ray data points from \sax\/ \citep{ds01}, and soft \g-ray data points from \integral\/ IBIS \citep{zls12} and \gro\/ COMPTEL \citep{mcconnell02}. Note that the spectrum below 20 keV represents a typical hard state spectrum of the source (which is absorbed by an intervening medium). Finally, black squares and arrows at high energies denote the detected fluxes and the derived upper limits from \fermi-LAT (30 MeV--0.3 TeV; MZC13) and MAGIC \citep[$>0.1$ TeV;][]{magic}. The green short-dashed curve corresponds to the stellar blackbody continuum. The dotted cyan curve shows the estimated unabsorbed accretion spectrum with (a) and without (b) a hybrid Comptonization tail \citep{pv09}. The red solid curves give the model jet-synchrotron spectra. The magenta dotted, cyan short-dashed, green long dashed, and blue solid curves illustrate the pair-absorbed SSC, BBC, XC, and the sum of the SSC+BBC+XC jet components, respectively. (a) The model 1 with soft electron injection spectrum ($p=2.5$), in which the MeV tail is due to hybrid Comptonization in the accretion flow (cyan dotted curve). (b) The model 2 with hard electron injection spectrum ($p=1.4$), in which the observed MeV tail is due to the jet synchrotron emission. 
} \label{spectra}
\end{figure*}

Upper limits on the flux from Cyg X-1 at photon energies $>30$ MeV and a detection of 0.1--10 GeV emission in the hard spectral state of Cyg X-1 have recently been reported using the \fermi\/ Large Area Telescope (LAT; \citealt*{mzc13}, hereafter MZC13). Although the detection has a limited statistical significance, it has been confirmed by the independent work of \citet{bodaghee13}, who found variable emission using the LAT data in the hard and intermediate states, but not in the soft state. The spectra and upper limits of MZC13 complement the previously known radio--to--hard X-ray spectra of Cyg X-1 in the hard state, e.g., the average one compiled in \citet*{zls12} and shown in Fig.\ \ref{spectra}. We apply to these data the jet model developed in \citet{z14}, hereafter Paper I. 

We take into account, in particular, the possibility that the MeV tail in the hard state of the source is due to the jet synchrotron emission, as implied by the recent claims of very strong linear polarization in that energy range \citep{l11,jourdain12}. We find that the combination of a high flux around MeV energies (if interpreted as synchrotron radiation) and a low flux above 30 MeV requires rather strong jet magnetic field (much above the equipartition level), necessary to reduce the strength of Compton scattering in high-energy \g-rays.

We note that the statistical significance of the result of \citet{l11} appears rather low, since the distribution of the azimuthal scattering angle presented in their fig.\ 2 is independent only up to $180\degr$. There is also a disagreement regarding the polarization in the 250--400 keV band, which was found weak and consistent with null by \citet{l11}, whereas \citet{jourdain12} found it to be $\simeq 50$ per cent. In addition, \citet{l11} found the polarization above 400 keV to be $67\pm 30$ per cent compared to the best fit at $>100$ per cent in the 370--850 keV band found by \citet[see figure\ 4 therein]{jourdain12}. The agreement of the polarized fraction with \citet{l11} at $76\pm 15$ per cent claimed by \citet{jourdain12} was obtained only by adding the two channels within the 230--850 keV range. The results regarding the 230--370 keV band are thus quantitatively different between the two papers. This could be, in principle, due to the somewhat different observation periods on which the two works are based, 2003--2009 for \citet{l11} vs.\ 2006--2009 for \citet{jourdain12}. 

Then, in the 370--850 keV band of \citet{jourdain12}, the minimum $\chi^2$ appears to be obtained at the polarization fraction of $\sim 150$ per cent. If we compare the $\chi^2$ not at this unphysical value but at $\sim 70$ per cent polarization (consistent with the stated fractions of \citealt{l11} and \citealt{jourdain12}) with the $\chi^2$ at null polarization, the resulting $\Delta\chi^2$ is $\la 5$ for 41500 degrees of freedom. Thus, the statistical significance of the presence of strong polarization at that channel appears weak. Therefore, we also consider models in which the jet does not account for the MeV tail.

In Section \ref{method}, we outline the method applied to model the data, and specify the adopted parameters of Cyg X-1. We extend the model presented in Paper I for an analytical treatment of the particle evolution beyond the energy dissipation region (see Appendix \ref{advection}). Sections \ref{soft}--\ref{gammamin} give results of the application of our model to the average hard-state spectrum of Cyg X-1. We present here a number of alternative models fitting the data. In Section \ref{magic_flare}, we present two models reproducing the spectrum of the TeV flare detected by MAGIC \citep{magic}. We discuss our results in Section \ref{discussion}, and give our conclusions in Section \ref{conclusions}.

\begin{table*}
\begin{center}
\caption{Parameters of Cyg X-1 adopted in this work.
}
\begin{tabular}{cccccccccccccccc}
\hline
$P$ & $M$ & $M_*$ & $r_*$ & $T_*$ & $i$ & $D$ & $\beta_{\rm j}$ & $\Theta_{\rm j}$ & $z_{\rm M}$ & $E_{\rm t0}$ & $F$(15 GHz)& $L_*$ & $a$ & $\Gamma_{\rm j}$ & $R_{\rm g}$\\
d & $\msun$ & $\msun$ & $\rsun$ & K & deg & kpc & & deg & cm & eV & mJy & erg s$^{-1}$ & cm & & cm\\
\hline
5.6 & 16 & 27 & 19 & $2.8\times 10^4$ & 29 & 1.86 & 0.6 & 2 & $10^{15}$ & 0.15 & 13 & $8\times 10^{38}$ & $3.2\times 10^{12}$ & 1.25 & $2.36\times 10^6$ \\
\hline
\end{tabular}
\end{center}
\label{cygx1}
\end{table*}

\setlength{\tabcolsep}{3.5pt}

\begin{table*}
\begin{minipage}{17.7cm}
\caption{The free parameters of the models and the derived quantities.}
\begin{tabular}{@{}lcccccccccccccccccc}
\hline
model & $p$ & $\gamma_{\rm m}$ & $B_0$ & $z_{\rm m}/ R_{\rm g}$ & $a/z_{\rm m}$ & $\gamma_{\rm t0}$ & $\gamma_{\rm b0}$ & $\lg(\beta_{\rm eq})$ & $\lg(\sigma_{\rm eq})$ & $\lg(P_{\rm e})$ & $\lg(P_{\rm i})$ & $\lg(P_B)$ & $\lg(P_{\rm inj})$ & $\lg(P_{\rm ad})$ & $\lg(P_{\rm S})$ & $\lg(P_{\rm BBC})$ &                                           $\lg(R_{\rm inj})$ & $\lg(R_{\rm e})$\\
&&&$\times 10^4$ &&&$@z_{\rm m}$ &$@z_{\rm m}$ &$@z_{\rm M}$ & $\leq $ &$@z_{\rm M}$ &$\geq $ &&&&&&& $\geq$\\
\hline
1 &  2.5  & 2   & 0.9   & 777 & 1760  & 29 & 78 & 1.2 & $-3.7$ & 34.7 & 36.6 & 33.5 & 35.7 & 35.6 & 34.0 & 33.6 & 40.9 & 40.0\\
1m &  2.5  & 300 & 7   & 341 & 4010  & 10 & 2.9 & $-0.42$ & $-0.60$ & 34.1 & 34.4 & 34.6 & 35.0 & 34.8 & 34.6 & 33.7 & 38.0 & 37.8 \\
2 &  1.4  & 2   & 50   & 285 & 4800  & 3.9 & 0.07 & $-1.6$ & 2.4 & 34.5 & 33.8 & 36.1 & 35.8 & 34.7 & 35.8 & 33.6 & 37.8 & 37.3\\
2a &  1.43  & 2   & 50   & 271 & 5060  & 3.9 & 0.07 & $-1.6$ & 3.6 & 34.5 & 33.9 & 36.1 & 35.8 & 34.7 & 35.8 & 33.7 & 38.0 & 37.3\\
2m &  1.5  & 300 & 100   & 167 & 8180 & 2.8 & 0.03 & $-1.7$ & 3.8 & 34.5 & 33.7 & 36.3 & 35.8 & 34.7 & 35.8 & 33.5 & 37.2 & 37.1\\
\hline
MZC-1 &  3.2 & 2   & 0.25 & 829 & 1650 & 55 & 950 & 3.8 & $-6.5$ & 36.3 & 38.3 & 32.4 &--& 37.1 & 33.8 & 34.2 & -- & 41.7 \\
MZC-2 &  2.3 & 2   & 4   & 1110 & 1230 & 14 & 2.8 & $-1.7$ & $-0.52$ & 33.4 & 34.9 & 35.1 &--& 34.2 & 35.8 & 33.1 & -- & 38.3\\
\hline
\end{tabular}
{\it Notes:} Parameters of the different models discussed in Sections \ref{soft}--\ref{gammamin} of this paper are compared with those of the models 2, 1 presented in MZC13 (denoted here as MZC-1 and MZC-2). In the models 1, 1m, and MZC-1 the MeV tail of Cyg X-1 is assumed to originate in the accretion flow, whereas in the models 2, 2a, 2m, and MZC-2 it is accounted by the jet synchrotron emission. The model 2a is a variant of 2 with the advection taken into account. The models 1m and 2m are variants of models 1 and 2, respectively, with high low-energy cutoff in the electron injection function. The first four listed quantities are the model free parameters, and the remaining ones are the model-derived parameters. The values of $\sigma_{\rm eq}$ and $P_B$ are for tangled magnetic field. Various components of the jet power provided here can be compared to the average hard-state bolometric accretion luminosity of Cyg X-1, $\lg(L_{\rm accr})\simeq 37.3$, and the Eddington luminosity, $\lg(L_{\rm E})\simeq 39.3$. The units of $B$, $P$ and $R$ are G, erg s$^{-1}$, s$^{-1}$, respectively.
\label{t:models}
\end{minipage}
\end{table*}

\section{The Model, Parameters, and Assumptions}
\label{method}

We use the same parameters of Cyg X-1 and its jet as those adopted by MZC13, which we list in Table \ref{cygx1} (note that the last four parameters are not independent). The most recent determination of the parameters of the binary is that of \citet{ziolkowski14}, who re-examined previous estimates of \citet{ziolkowski05}, \citet{cn09} and \citet{orosz11}. Different than \citet{orosz11}, \citet{ziolkowski14} took into account that Cyg X-1 is still in the stable, core H-burning, phase, as evidenced by the observed stability of the period, $P$, and of the average X-ray luminosity \citep[see discussions in][]{ziolkowski05,ziolkowski14}. This implies a tight correlation between the mass and the luminosity, which is violated by the best fit of \citet{orosz11}. Taking it into account, we adopt the black hole mass, $M$, the donor mass, $M_*$, radius, $r_*$, the effective temperature, $T_*$, and the binary inclination, $i$, as given in Table \ref{cygx1}. The distance, $D$, is from \citet{reid11}, the jet velocity, $\beta_{\rm j}$ (assumed to be constant along the jet), is based on \citet{stirling01}, \citet{gleissner04} and \citet*{mbf09}, and the opening angle, $\Theta_{\rm j}$, is the upper limit of \citet{stirling01}. These parameters correspond to the stellar luminosity, $L_*$, the separation between the components, $a$, the jet bulk Lorentz factor, $\Gamma_{\rm j}$, and the gravitational radius, $R_{\rm g}$, as given in Table \ref{cygx1}.

We assume the jet is perpendicular to the binary plane. The radio structure seen by Very Large Array (VLA) and Very Long Baseline Array (VLBA) at 8.4 GHz extends up to $z_{\rm M}\sim 10^{15}$ cm \citep{stirling01,rushton09,rushton11}. We assume this value as the jet height, which corresponds to $z_{\rm M}\simeq 300 a$, and which also approximately equals the height of the dissipation region in the internal shock model for the hard state of black-hole binaries of \citet{malzac13}. The entire emitting jet becomes optically thin above the turnover energy, $E_{\rm t0}$. The adopted value of $E_{\rm t0}$ (Table \ref{cygx1}) is based on the results of \citet{rahoui11}. The flux in the partially self-absorbed region is normalized to the average hard-state flux at 15 GHz (Table \ref{cygx1}). 

Fig.\ \ref{spectra} includes the same average hard-state broad-band spectrum of Cyg X-1 as that of \citet{mzc13} except that we now show also the average fluxes at 235 MHz and 610 MHz based on \citet{pandey07}. Given that one of their 610-MHz measurements has an error much smaller than all other ones, the weighted average is strongly dominated by that single measurement even if we include two non-detections (on MJD 53104, 53127) as null fluxes with the errors equal to the upper limits. Therefore, we give the unweighted averages of the detections only (MJD 52891, 53102, 53107), yielding $5.8\pm 3.0$ mJy and $8.3\pm 1.3$ mJy, at 235 MHz and 610 MHz respectively. 

Theoretical spectra of the jet emission are calculated using the model developed in Paper I. We take into account all relevant radiative processes. These are synchrotron radiation, synchrotron self-Compton (SSC), and Compton upscattering of stellar blackbody photons (BBC) and of the accretion flow photons (XC). The accretion flow photons are described with a disc and hot inner flow model (see \citealt{yn14} and references therein). The accretion disc is assumed to extend between $r_{\rm in}=5\times 10^7$ cm and $r_{\rm out}=2\times 10^{11}$ cm (the exact values of which influence only little our results), and to have the maximum temperature of $kT_{\rm in}=150$ eV with the colour correction of $f_{\rm c}=2$, based on \citet{st95}. This disc soft X-ray spectrum agrees well with the observed spectrum of \citet{ds01}, which is typical for the hard state. At high energies, we use the hybrid Comptonization fit to the \integral\/ data of \citet{zls12}. The resulting model spectrum is shown in Fig.\ \ref{spectra}. Its total luminosity is $L_{\rm accr}\simeq 2.0\times 10^{37}$ erg s$^{-1}$.

We use the same notation as Paper I (see Section 2 therein for details). The height along the jet is expressed in units of $z_{\rm m}$, $\xi\equiv z/z_{\rm m}$, where $z_{\rm m}$ marks the onset of the energy dissipation within the outflow. The magnetic field strength is $B=B_0(z/z_{\rm m})^{-1}$, and the jet is assumed to be conical with the opening angle of $\Theta_{\rm j}$. The minimum electron Lorentz factor, $\gamma$, down to which we calculate the steady-state distribution, is assumed to be $\gamma_0=2$. The electrons are accelerated in the dissipation region, $z_{\rm m}\leq z\leq z_{\rm M}$, and injected with a power-law rate above a minimum Lorentz factor of $\gamma_{\rm m}$, 
\begin{equation}
Q(\gamma,z) = \cases{0,&$\gamma<\gamma_{\rm m}$ or $z_{\rm m}>z>z_{\rm M}$;\cr
Q_0(z/z_{\rm m})^{-3} \gamma^{-p}g_{\rm cut}(\gamma, \gamma_{\rm M}), &$\gamma\geq \gamma_{\rm m}$ and $z_{\rm m}\leq z\leq z_{\rm M}$,}
\label{Q_inj}
\end{equation}
where $g_{\rm cut}(\gamma,\gamma_{\rm M})$ describes the high-energy cutoff (see Paper I). Given theoretical uncertainties in the determination of the exact spectral shape of a cut-off in the injection particle distribution $Q(\gamma, z)$, instead of specifying the $g_{\rm cut}(\gamma, \gamma_{\rm M})$ function, below we simply assume the super-exponential form of the high-energy cut-off in the corresponding steady-state electron distribution $N(\gamma, z)$, namely
\begin{equation}
f_{\rm cut}(\gamma,\gamma_{\rm M})=\exp\left[-(\gamma/\gamma_{\rm M})^2\right],
\label{fcut}
\end{equation}
where $\gamma_{\rm M}(B_0,z,\eta_{\rm acc})\propto z^{1/2}$ is given by equations (6) and (5) of Paper I. With the adopted $\delta$-function approximation in the synchrotron emissivity, $E/(m_{\rm e} c^2) =(B/B_{\rm cr})\gamma^2$, the cutoff in the model synchrotron spectra is then exponential, $\propto \exp(-E/E_{\rm M})$, where $E_{\rm M}$ is the corresponding photon e-folding energy. 

Paper I presented two methods of calculating $N(\gamma,z)$. In the simpler method, the electron distribution is solved locally including the effect of self-absorption, and electron advection along the jet is neglected, This yields a relatively simple formula for $N(\gamma,z)$,
\begin{equation}
N(\gamma,z)={-1\over \dot{\gamma}(\gamma, z)} \int_\gamma^{\infty}{\rm d}\gamma'' Q(\gamma'',z).
\label{N_cool}
\end{equation}
The effect of the synchrotron self-absorption, which makes $\dot{\gamma}(\gamma, z)$ itself dependent on $N(\gamma,z)$,  can be taken into account using the approximation of equation (12) of Paper I. The equation for $N$ becomes then a quadratic one, see equation (46) of Paper I.
 
In the other method, electron advection along the jet is taken into account via the continuity equation solved in both electron energy and length along the jet,
\begin{equation}
{\Gamma_{\rm j}\beta_{\rm j}c\over z^2} {\partial\over \partial z}\left[z^2 N(\gamma,z)\right]+{\partial\over \partial \gamma}\left[\dot \gamma(\gamma,z) N(\gamma,z)\right]=Q(\gamma,z),  
\label{ndot}
\end{equation}
where $\dot\gamma$ is the total electron energy loss rate. If Klein-Nishina (KN) effects (important in Cyg X-1) are taken into account, this transport equation has to be solved numerically. Also, in such a case there is no simple way to account for self-absorption effects. Given that calculating Compton-scattered spectra averaged over the binary phase and jet length involves additional multi-dimensional integrals (see Paper I), using a numerical solution for $N(\gamma,z)$ increases substantially the complexity of the calculations and the computation time. 

However, it has been shown in Paper I that the two solutions are very close to each other for the particular form of $Q(\gamma,z)$ assumed here (equation \ref{Q_inj}), i.e., for constant monochromatic power injected per $\ln z$. For adiabatic losses only, the two solutions coincide except around $z=z_{\rm m}$, at which height the advective solution is assumed to have the null boundary condition. When radiative losses are added, the two solutions converge even faster (see fig.\ 4 in Paper I). Thus, we use in the present paper the local solution including continuous electron injection and cooling, equation (\ref{N_cool}), within the dissipation region. Still, in order to quantify the effect of advection, we also discuss the analytical advective solution for the case of dominant synchrotron losses (equation 40 in Paper I) up to $z_{\rm M}$. We compare it with our model 2 (Section \ref{hard}), for which synchrotron losses dominate. 

On the other hand, the electron advection becomes crucial at $z>z_{\rm M}$, i.e., above the dissipation region, since the only relativistic electrons there are those advected from $z\leq z_{\rm M}$. As we found out in our calculations, relativistic electrons at $z = z_{\rm M}$ may still carry a significant fraction of the jet power, $P_{\rm e}$, comparable to that carried by the other plasma constituents (see Table \ref{t:models}). This power is then radiatively and adiabatically lost at $z>z_{\rm M}$. In order to treat that region, we extend the model described in Paper I for the analytical treatment of the particle evolution beyond the dissipation zone, as presented in Appendix \ref{advection} below. We obtain a very simple formula for $N(\gamma,z>z_{\rm M})$, given by equations (\ref{N_adv}--\ref{gamma_max}).

After specifying the fixed system parameters as given in Table \ref{cygx1}, the six free parameters of a model are $\eta_{\rm acc}$, $Q_0$, $p$, $z_{\rm m}$, $B_0$, and $\gamma_{\rm m}$, which can be determined using the spectrum observed from Cyg X-1, augmented by theoretical considerations. The dimensionless factor $\eta_{\rm acc}$ determines $E_{\rm M}$. We assume here $\eta_{\rm acc}=0.008$, which reproduces the cutoff of the MeV tail of Cyg X-1. We choose the value of either $\gamma_{\rm m}=2$ or 300, which corresponds to a usual single power-law injection or that with a low-energy cutoff due to equipartition with ion energy (see the discussion in Paper I). The remaining four parameters can be determined by four other observables. Here we choose them as the flux at 15 GHz, the turnover energy (Table \ref{cygx1}), and the fluxes at 1 MeV and 1 GeV. The main observational uncertainty is the jet flux at 1 MeV. As discussed in Section \ref{intro}, the observed MeV flux may or may not be due to synchrotron emission of the jet. If we assume it does, then we specify $F(1\,{\rm MeV})$ uniquely. If we attribute the MeV tail to another process (most likely hybrid Comptonization in the accretion flow), then we are free to choose $F(1\,{\rm MeV})$, as long as it is well below the observed flux. We then follow the solution method outlined in section 6 of Paper I.

In Table \ref{t:models}, we list the obtained values of the free parameters except for $Q_0$, for which we provide instead both the total rate and the power in the injected electrons, $R_{\rm inj}$ and $P_{\rm inj}$, given by equations (54) and (55) of Paper I, respectively. Table \ref{t:models} also gives the Lorentz factor of electrons providing the dominant contribution to the emission (at $z_{\rm m}$) at $E_{\rm t0}$, $\gamma_{\rm t0}$, and the Lorentz factor at which $N(\gamma,z_{\rm m})$ steepens due to radiative cooling, $\gamma_{\rm b0}$. The electron equipartition parameter, $\beta_{\rm eq}$, depends on the pressure of relativistic electrons, which evolves along the outflow due to radiative cooling. Radiative cooling influences also the power carried by relativistic electrons, $P_{\rm e}$. Thus, we give both quantities at the maximum height of the dissipation region, where they are maximized. The magnetization parameter, $\sigma_{\rm eq}$, is the ratio of the magnetic enthalpy to the particle enthalpy. The latter has a major contribution from ions, while in our model only a lower limit on the ion density is obtained from the density of relativistic electrons. The ion enthalpy is also proportional to the ion contribution to the jet power, $P_{\rm i}$. Thus, $\sigma_{\rm eq}$ and $P_{\rm i}$ parameters in Table \ref{t:models} correspond to the upper and lower limits, respectively. Also, the flux of all the electrons in the jet, $R_{\rm e}$, may be contributed substantially by cold electrons, and therefore the given value is a lower limit only. Table \ref{t:models} also gives the jet power in the magnetic field, $P_B$, the components of the radiative power in the synchrotron, $P_{\rm S}$, and the BBC process, $P_{\rm BBC}$, and the power lost adiabatically, $P_{\rm ad}$. 

MZC13 have presented jet models of Cyg X-1 neglecting cooling and advection, and instead assuming the electron steady-state distribution to be a single power law with an exponential cutoff; such an approach is commonly used in the literature \citep*[see, e.g.,][]{brp06}. For completeness, we also give the parameters of the models of MZC13 in Table \ref{t:models}.

We note that an inner part of the counterjet, $z<r_{\rm out}/\tan i\sim 10^{12}$ cm, is obscured by the accretion disc. Also, the counterjet emission is de-boosted. Thus, we do not include the synchrotron emission from the counterjet in our model spectra. On the other hand, we take into account the counterjet contribution to the Compton spectra, which is enchanced due to the Compton anisotropy, favouring large-angle scattering. Also, we include the counterjet in calculating components of the system power.

\section{Electron and Emission Spectra of Cyg X-1}
\label{models}

\subsection{Soft injection spectrum}
\label{soft}

We first consider the case with the standard, soft electron-injection spectrum, $p=2.5$ (corresponding to the spectral index of the uncooled optically-thin synchrotron emission of $\alpha=0.75$). With this, we calculate the model spectra (hereafter model 1) shown in Fig.\ \ref{spectra}(a), with the parameters given in Table \ref{t:models}. The obtained value of $z_{\rm m}$ corresponds to the maximum dimensionless height of $\xi_{\rm M}\equiv z_{\rm M}/z_{\rm m}\simeq 5.4\times 10^5$. The spectrum around the turnover energy at the jet base is due to the uncooled electrons. In this model, the MeV tail is not produced by the jet, but instead has to be due to a different process, most likely hybrid Comptonization in the accretion flow \citep*{mcconnell02,pv09,mb09,vpv11,vpv13}. 

We find that virtually all the observed emission (including radio) originates from $z<z_{\rm M}$. Although the electrons still carry a significant power at $z_{\rm M}$, $P_{\rm e}\sim 0.1 P_{\rm inj}$ in Table \ref{t:models}, that power is lost at $z>z_{\rm M}$ mostly adiabatically, with only $\sim 10^{-3}$ radiated away via the BBC process, which is in part due to the softness of the injected electron distribution.

The electron equipartition parameter is $\sim 10$, but since electron energy density is dominated by the lowest energy particles, we can reach $\beta_{\rm eq}\sim 1$ by a moderate increase in $\gamma_{\rm m}$ (which is very weakly constrained by the observed spectrum). Alternatively, we can obtain $\beta_{\rm eq}\sim 1$ by a slight decrease of $p$, but then the stronger magnetic field required to reproduce $F(1\,{\rm GeV})$ would result in $\gamma_{\rm b0}>\gamma_{\rm t0}$. This inequality would imply that the synchrotron emission at the turnover is dominated by the cooled electrons, and the emission continuum has no cooling break in its optically-thin part (which, however, is not constrained by observations). The magnetization parameter is very low, and the flow dynamics is dominated by ions. This model accounts well for the 0.1--0.3 and 1--10 GeV fluxes measured with the LAT. 

\begin{figure}
\centerline{\includegraphics[width=6.8cm]{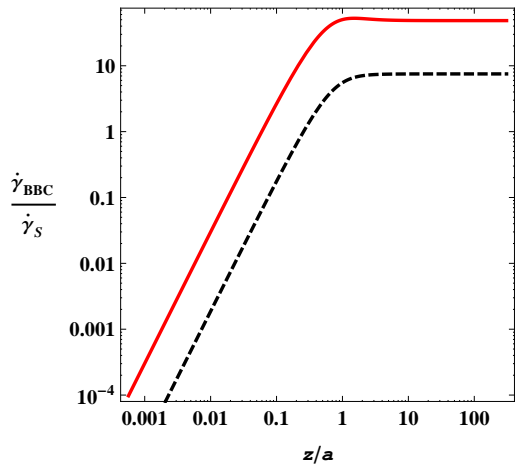}} 
\caption{The cooling rate ratio $\dot\gamma_{\rm BBC}/\dot\gamma_{\rm S}$ as a function of the jet height, $z/a$, in the Thomson limit and for the optically-thin synchrotron (i.e., for $\gamma \la 10^4$), and in the KN regime (at $\gamma=10^5$), shown by the red solid and black dashed curves, respectively,  for the model 1. In the KN case, this ratio is a function of both $z$ and $\gamma$. The plotted range of $z/a$ corresponds to $z$ between $z_{\rm m}$ and $z_{\rm M}$.
} \label{gdotbb}
\end{figure}

\begin{figure}
\centerline{\includegraphics[width=6.8cm]{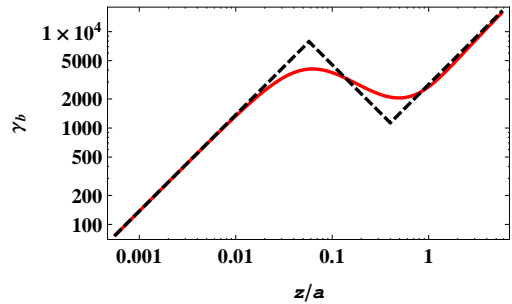}} 
\caption{The break electron Lorentz factor, $\gamma_{\rm b}$, vs.\ $z/a$ within the validity of the Thomson approximation for the parameters of the model 1. The red solid and black dashed curves give the accurate dependence of equation (49) and the approximation of equation (50) of Paper I, respectively.
} \label{gamma_break}
\end{figure}

\begin{figure}
\centerline{\includegraphics[width=7.2cm]{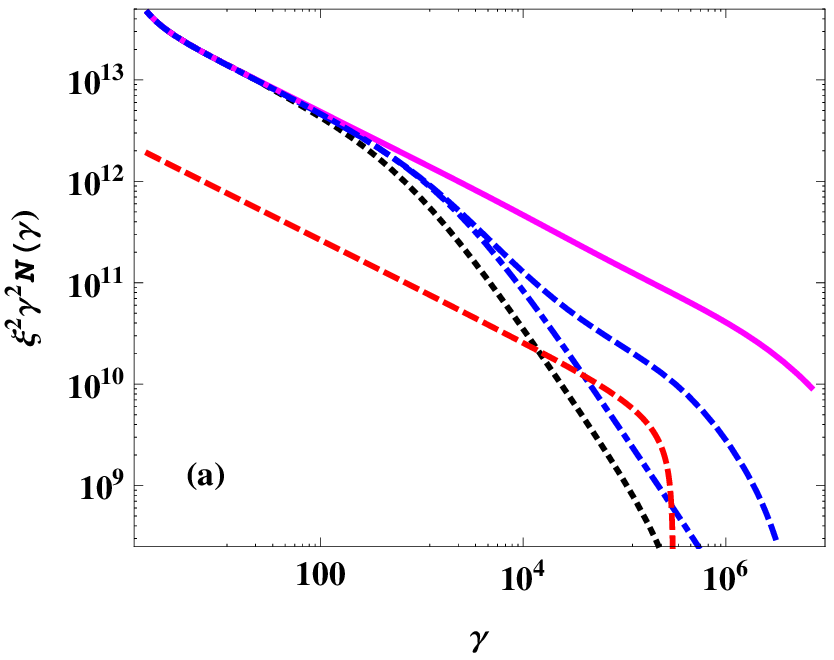}} 
\centerline{\includegraphics[width=7.2cm]{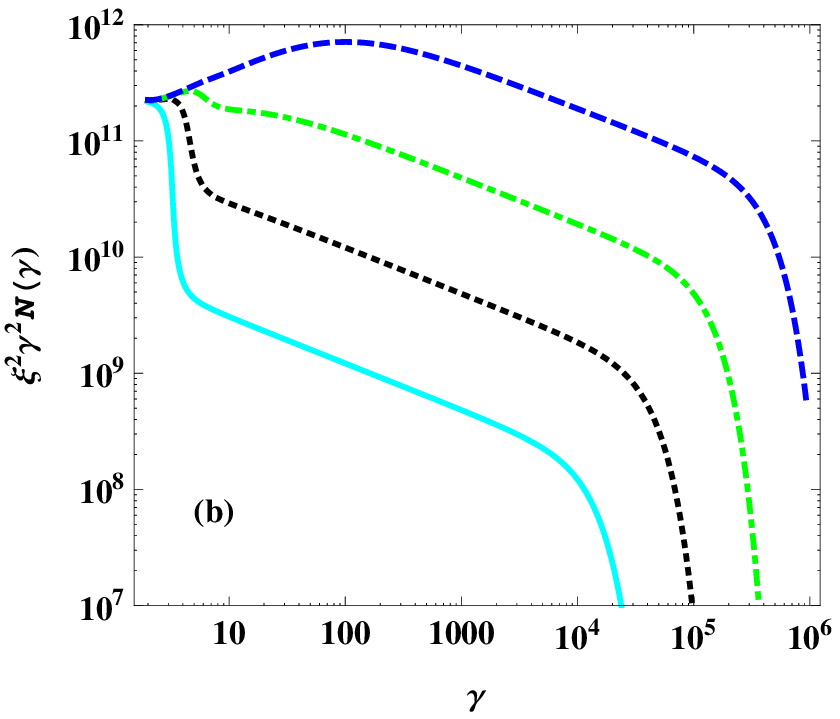}} 
\caption{Self-consistent electron distributions for (a) model 1. The black dotted, blue dashed, magenta solid and red dashed curves correspond to $\xi=10$, $10^3$, $10^5$, $10^7$ ($\simeq 18\xi_{\rm M}$), respectively. The last $N(\gamma,\xi)$ is due to advection of the electrons from the acceleration region. The blue dot-dashed curve shows the distribution in the Thomson approximation at $\xi=10^3$ ($\sim 0.5 a$), which illustrates the importance of KN effect around this height. (b) The model 2. The cyan solid, black dotted, green dot-dashed, and blue dashed curves correspond to $\xi=1$, $10^1$, $10^2$, $10^3$, respectively. We see a pronounced effect of the synchrotron self-absorption at low $\gamma$ and $\xi$, strongly reducing the synchrotron cooling and increasing $N(\gamma,z)$ at the lowest $\gamma$ to the values corresponding to the adiabatic cooling only. If radiative cooling were neglected altogether, the $\xi^2\gamma^2 N(\gamma,z)$ distributions would be independent of height for all models.
} \label{N_gamma}
\end{figure}

\begin{figure}
\centerline{\includegraphics[width=7.2cm]{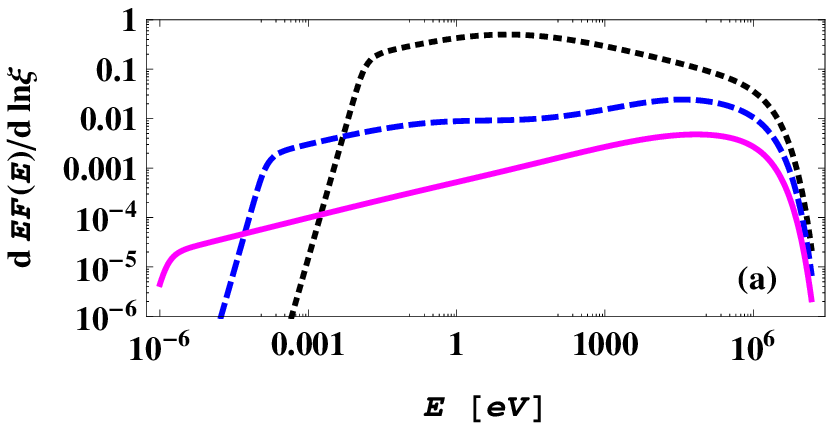}} 
\centerline{\includegraphics[width=7.2cm]{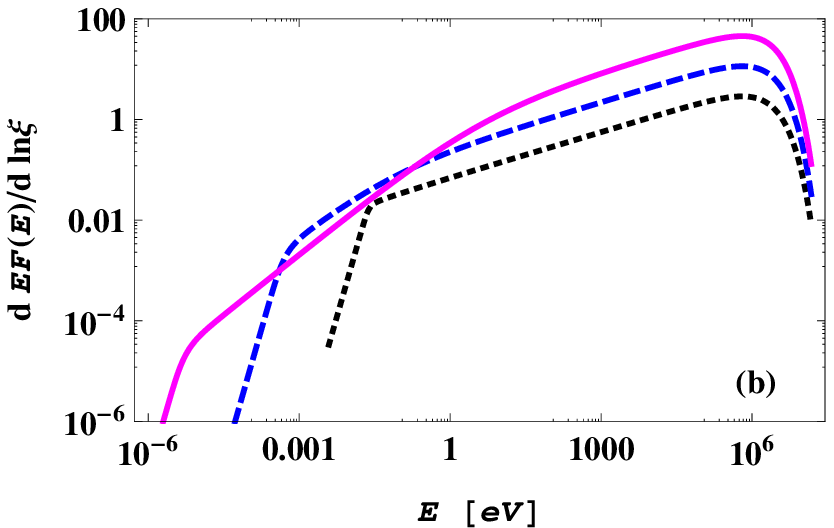}} 
\centerline{\includegraphics[width=7.2cm]{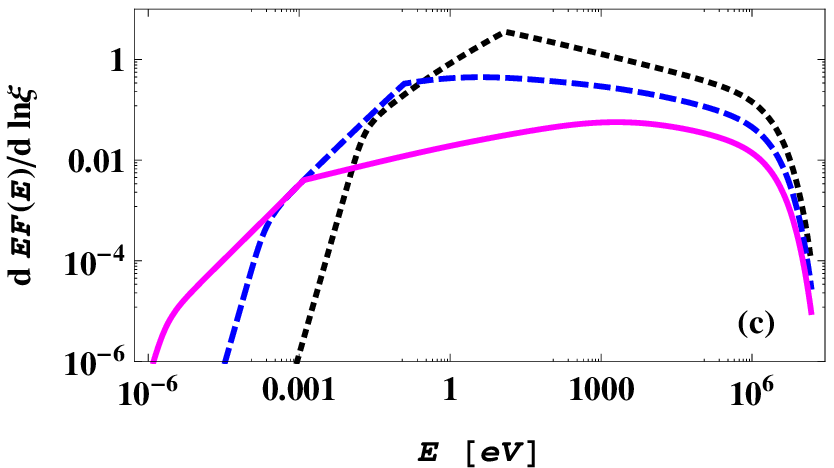}} 
\caption{Contributions to the total synchrotron spectrum of the model (a) 1, (b) 2, and (c) 1m, shown in Figs.\ \ref{spectra}(a--b) and \ref{spectragmin}, from the logarithmic height intervals at $\xi=10$, $10^3$, $10^5$, denoted by the black dotted, blue dashed, and magenta solid curves, respectively. 
} \label{syn_emissivity}
\end{figure}

\begin{figure}
\centerline{\includegraphics[width=7.2cm]{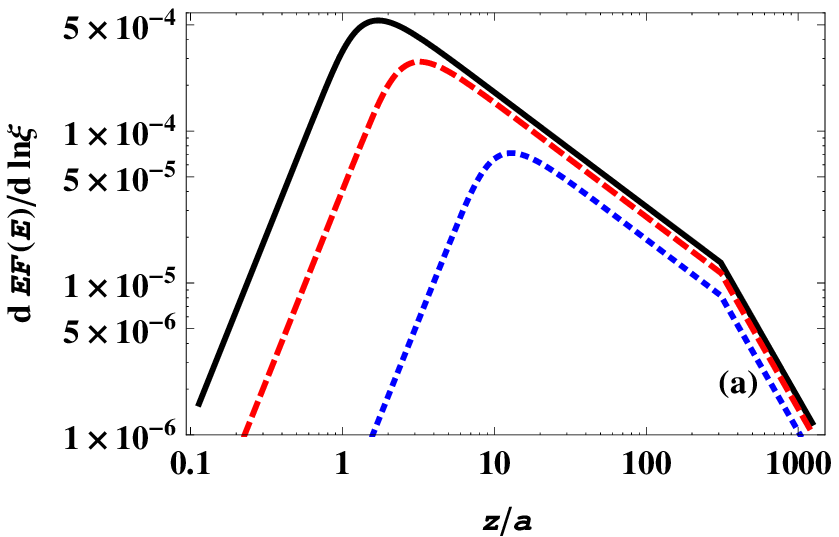}} 
\centerline{\includegraphics[width=7.2cm]{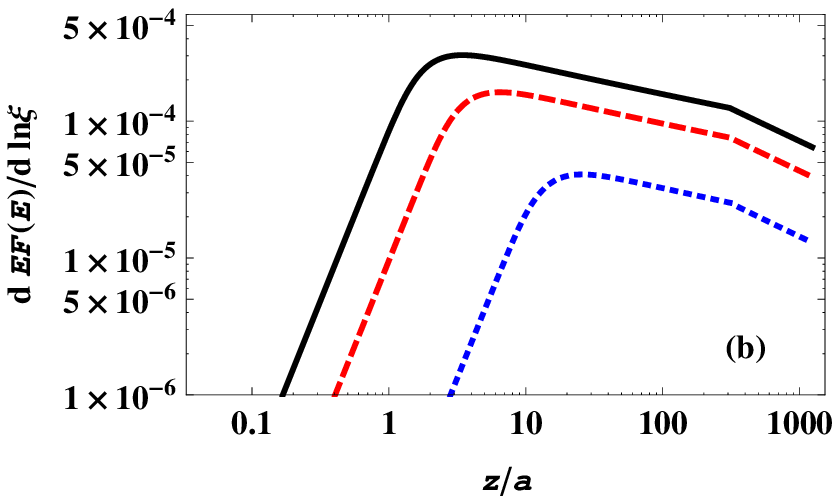}} 
\caption{Vertical profiles of the contribution to the synchrotron flux per unit $\ln \xi$ at 2 GHz, 8 GHz and 15 GHz, shown by the blue dotted, red dashed and black solid curves, respectively, for the model (a) 1 and (b) 2. We can see that the observed flux at a given frequency has a higher contribution from the optically-thin part (at high $z$) than from the optically-thick part (at low $z$). The kinks correspond to $z_{\rm M}$, and the emission beyond it is from advected electrons.
} \label{syn_z_profile}
\end{figure}

\begin{figure}
\centerline{\includegraphics[width=7.2cm]{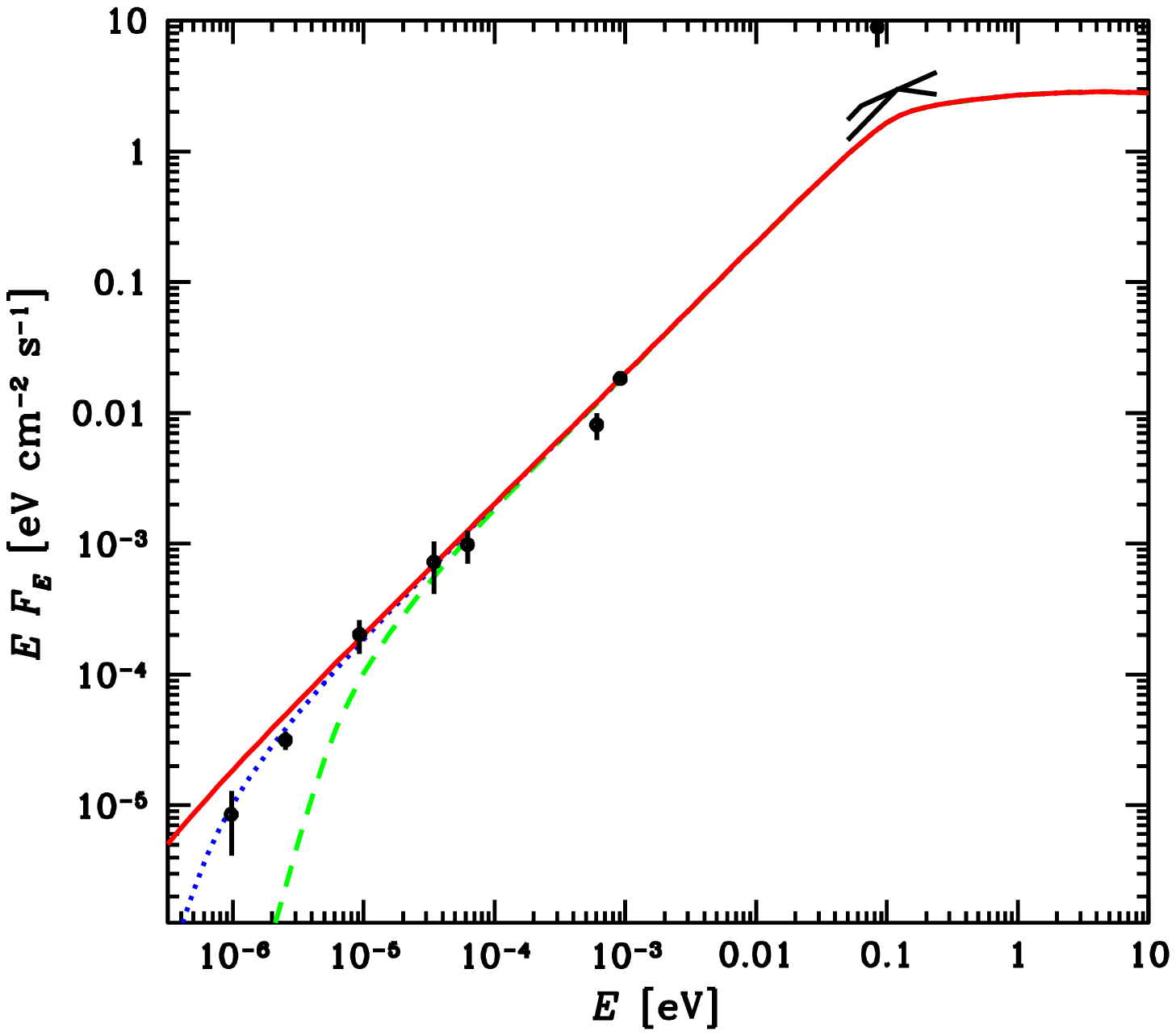}} 
\caption{The dependence of the synchrotron spectrum of the model 1 (shown in Fig.\ \ref{spectra}a) on the total jet height, for $\xi_{\rm M}\equiv z_{\rm M}/z_{\rm m}= 3\times 10^4$ ($\simeq 5.5\times 10^{13}$ cm), $3\times 10^5$ ($\simeq 5.5\times 10^{14}$ cm) and $3\times 10^6$ ($\simeq 5.5\times 10^{15}$ cm), shown by the green dashed, blue dotted, and red solid curves, respectively. The radio and mm data are also shown.
} \label{spec_zmax}
\end{figure}

In the model considered, the energy loss rate due to the SSC process, $\dot\gamma_{\rm SSC}$, is always much smaller compared to the total cooling rate, $\dot\gamma$, and thus this process is negligible for determination of the steady-state electron distribution, $N(\gamma,z)$. Similarly, the loss rate due to scattering of accretion photons, $\dot\gamma_{\rm XC}$ is significantly below the total loss rate. Specifically, the global maximum of $\dot\gamma_{\rm XC}/\dot\gamma\simeq 0.13$ is reached at $z=z_{\rm m}$ for $\gamma\simeq 120$. These results are consistent with the spectral components due to the SSC and XC processes being also of minor importance compared to the synchrotron and BBC spectra (see Fig.\ \ref{spectra}).

\begin{figure}
\centerline{\includegraphics[width=7.2cm]{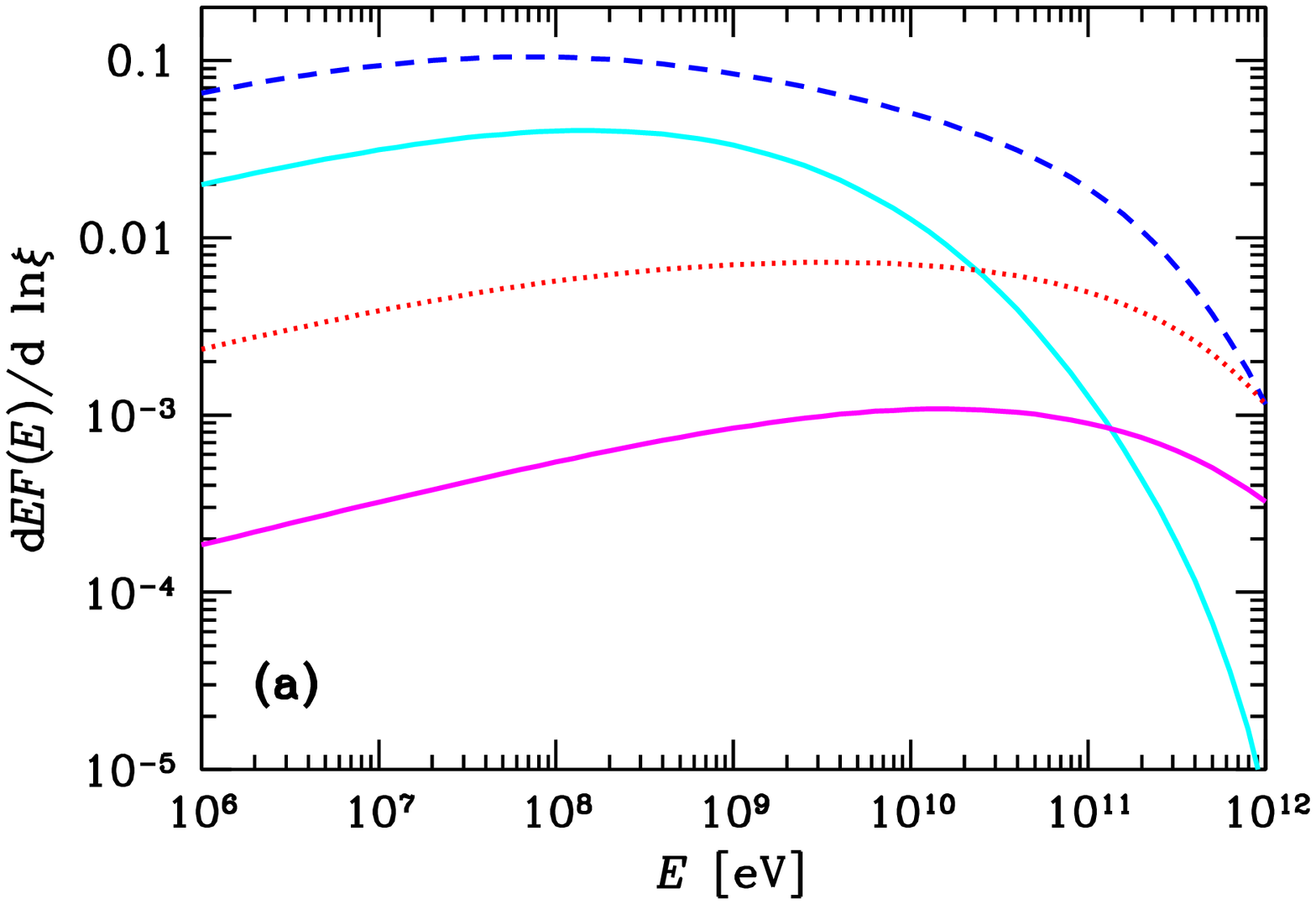}} 
\centerline{\includegraphics[width=7.2cm]{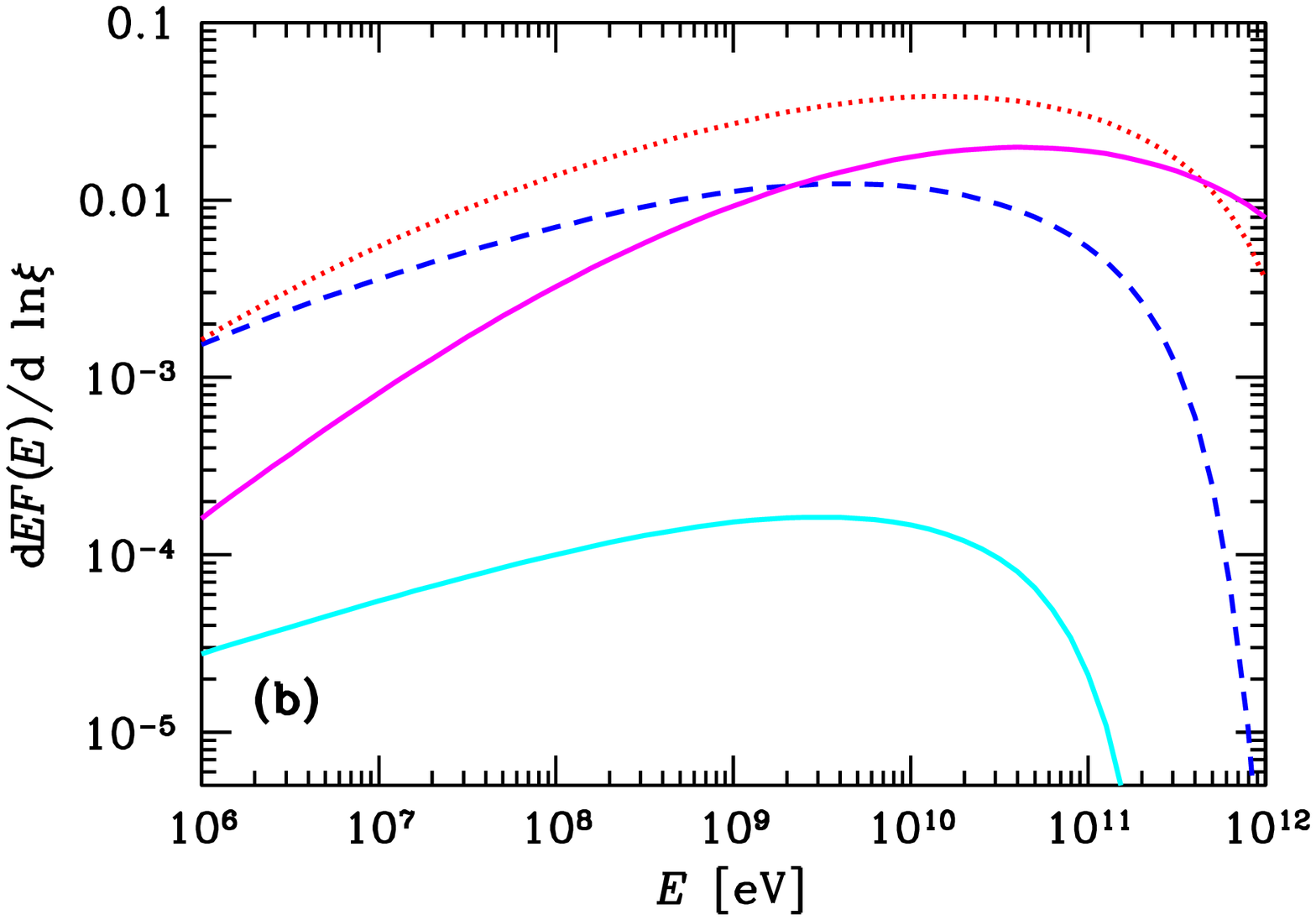}} 
\caption{Contributions to the total BBC spectrum (before pair absorption) of the model (a) 1, (b) 2, shown in Fig.\ \ref{spectra}, from the logarithmic height intervals at $\xi=10^2$, $10^3$, $10^4$, $10^5$, shown by the cyan solid, blue dashed, red dotted and magenta solid curves, respectively.
} \label{bc_emissivity}
\end{figure}

\begin{figure}
\centerline{\includegraphics[width=7.2cm]{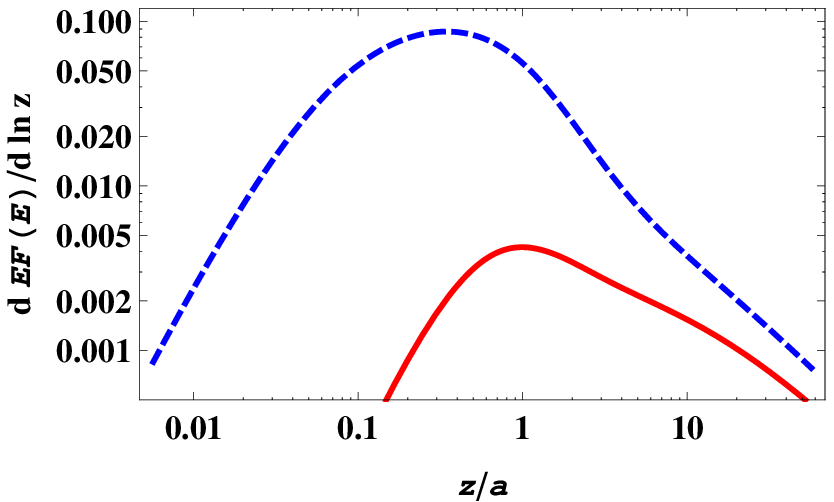}} 
\caption{Profiles of the emitted (i.e., before pair absorption) BBC flux per $\ln z$ for the model 1. The blue dashed and red solid curves are for $E=1$ GeV and 0.5 TeV, respectively. 
} \label{bc_profile}
\end{figure}

The solid curve in Fig.\ \ref{gdotbb} shows the ratio $\dot\gamma_{\rm BBC}/\dot\gamma_{\rm S}$ as a function of $z/a$ for scattering in the Thomson limit (where this ratio is independent of $\gamma$). The ratio is initially increasing $\propto z^2$ since the flux of the stellar radiation is approximately constant along the jet for $z\ll a$ whereas the magnetic field is decreasing. At $z\sim 0.05a$, the BBC losses start to dominate the synchrotron ones. Then, for $z\ga a$, the stellar radiation becomes diluted approximately as $z^{-2}$, which is the same dependence as for the synchrotron cooling. Thus, the $\dot\gamma_{\rm BBC}/\dot\gamma_{\rm S}$ ratio becomes a constant. The dashed curve shows the same ratio in the KN regime calculated for $\gamma=10^5$. The ratio is now lower, due to the KN reduction of $\dot\gamma_{\rm BBC}$, but the qualitative dependence on $z$ is the same. Fig.\ \ref{gamma_break} shows the position of the cooling break Lorentz factor, $\gamma_{\rm b}$ (see Paper I for details), along the jet (for $\gamma\leq 10^4$, up to which the Thomson limit applies). We see an approximately flat/inverted part in the range of $z\sim (0.03$--$1)a$, which is due to the dominance of the undiluted blackbody cooling in that region.

The steady-state electron distribution at four values of $\xi$ is shown in Fig.\ \ref{N_gamma}(a). The distribution at $\xi=10$ is dominated by adiabatic cooling at $\gamma \la 10^2$ and by synchrotron cooling at $\gamma \ga 10^3$. At $\xi=10^3$, we see first the cooling break due to the BBC process, and then a strong increase of $N(\gamma,z)$ with respect to that of the Thomson approximation (shown by the dot-dashed curve) at $\gamma\ga 10^4$, which is due to the KN reduction of the BBC cooling. The KN effect is maximized around this $\xi$ because the synchrotron and adiabatic cooling dominate at lower and higher values of $\xi$, respectively. At $\xi=10^5$, the cooling is mostly adiabatic. The slight bent at the lowest values of $\gamma$ is due to the adiabatic loss being linear in the momentum, $\beta\gamma$, rather than $\gamma$, equation (8) in Paper I. The distribution at $\xi=10^7$ is at $\simeq 18 z_{\rm M}$, and  thus consists only of electrons advected from the acceleration region and cooled on the way. 

Fig.\ \ref{syn_emissivity}(a) shows the contributions to the total synchrotron spectrum from logarithmic height intervals at three values of $\xi$. The superposition of the partially self-absorbed parts results in the power-law slope of the radio continuum of $\alpha=0$ below the turnover energy. The bulk of the observed optically thin synchrotron emission is produced around the vicinity of the jet base. Note that the slope of the optically-thin emission changes from mostly cooled one at the jet base to fully uncooled at large heights. The hardening seen at $E\ga 30$ keV in the spectrum at $\xi=10^3$ is due to the KN reduction of the Compton cooling of the emitting electrons. 

Fig.\ \ref{syn_z_profile}(a) shows the corresponding vertical profiles of the synchrotron flux at three radio frequencies, which are within the partially self-absorbed part of the spectrum. The dependencies of the fully self-absorbed and optically-thin parts are $\propto z^{5/2}$ and $\propto z^{(1-p)/2}$ at $z<z_{\rm M}$, see equation (71) of Paper I. Thus, the optically-thin contribution, emitted by uncooled electrons, dominates the total flux. The kinks in the profiles correspond to $z_{\rm M}$, and the emission beyond $z_{\rm M}$ is due to electrons advected from the region of $z\leq z_{\rm M}$. We see that the maximum of 15 GHz emission occurs at $z\simeq 1.7 a$. This is in a qualitative agreement with the observed strong orbital modulation at that frequency, which is well fitted by wind absorption of the jet emission from $z\sim a$ \citep{zdz12}. We also see the dependence of the height corresponding to the highest contribution of $z\propto \nu^{-1}$ \citep{bk79}. 

Fig.\ \ref{spec_zmax} shows the dependence of the jet synchrotron spectrum on $\xi_{\rm M}$. There is virtually no effect of $\xi_{\rm M}$ changing from $3\times 10^4$ to $3\times 10^6$ on the spectrum at $E\ga 10^{-3}$ eV. However, the lowest $\xi_{\rm M}$ gives the spectrum in the GHz range much below the radio data. Given that the transition from the optically thick to optically thin regimes take place approximately at $\nu\propto z^{-1}$ \citet{bk79}, the ratio $z_{\rm M}/z_{\rm m}$ has to be larger than the ratio of the highest to lowest energies in the optically-thick, $\alpha\simeq 0$, part of the spectrum. The best fit to the data appears to be provided by the middle case. Note that the two data points at the lowest energies are based on only three measurements each, see Section \ref{method}, making their averages relatively uncertain.

\begin{figure}
\centerline{\includegraphics[width=7.2cm]{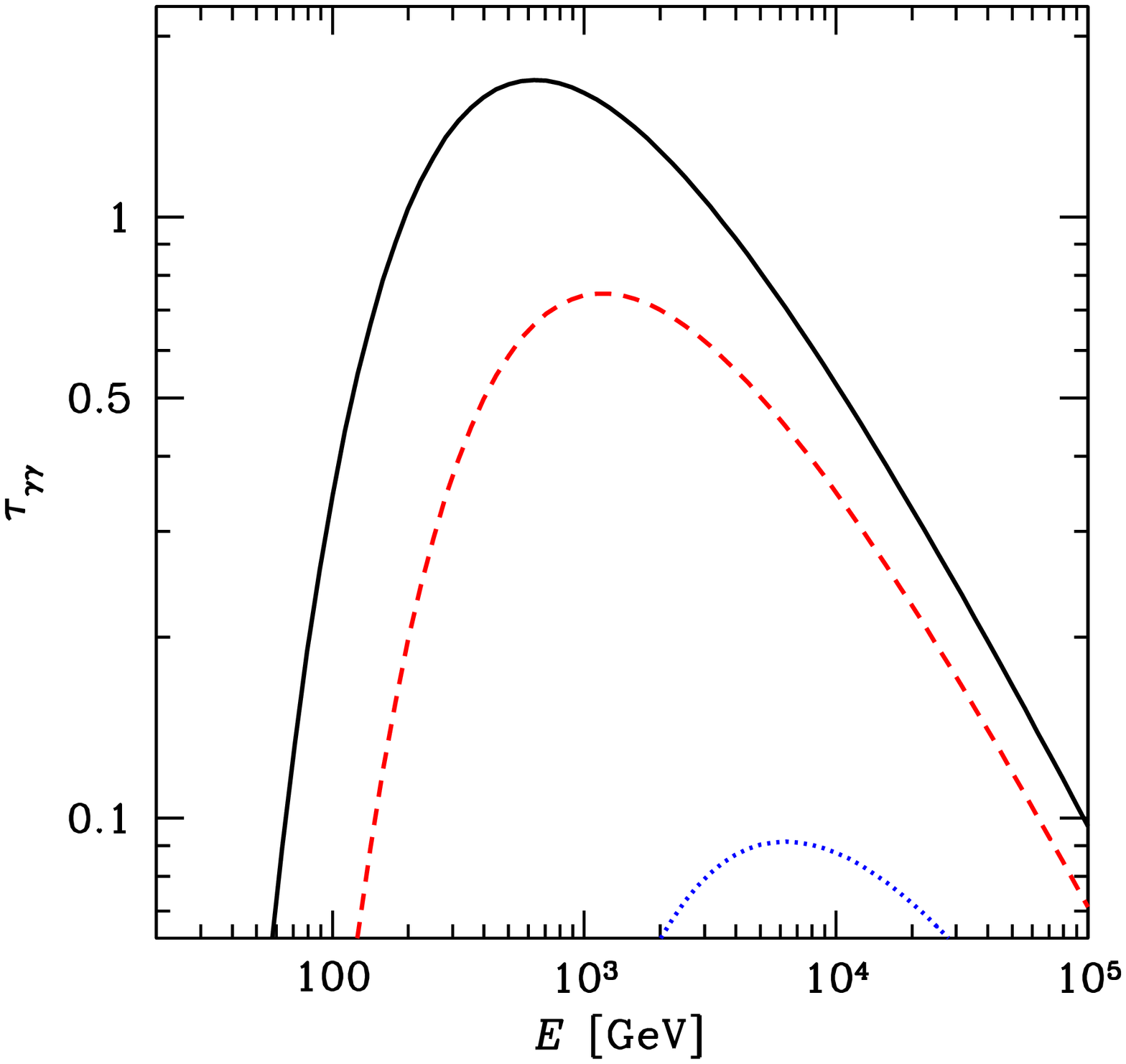}} 
\caption{The optical depth to pair production from the jet at the height of $3.2\times 10^{12}$ cm ($\simeq a$; at which the 0.5 TeV BBC emission is maximized) towards the observer (at $i=29\degr$) for the orbital phase of $\phi_{\rm b}=0$ (the superior conjunction), $\pm\upi/2$ and $\upi$ (the inferior conjunction), shown by the black solid, red dashed and blue dotted curves, respectively. 
} \label{taugg}
\end{figure}

\begin{figure}
\centerline{\includegraphics[width=7.2cm]{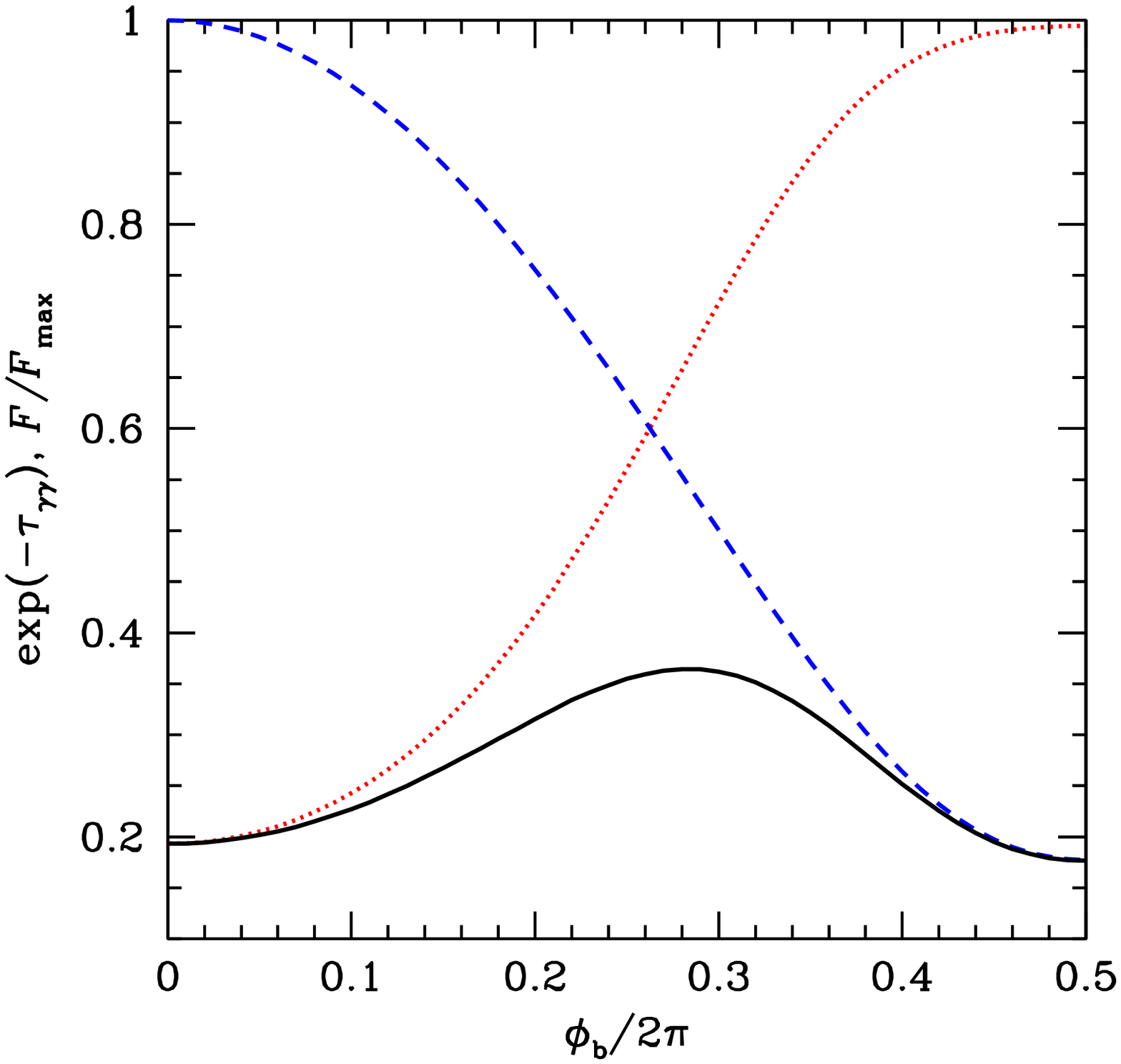}} 
\caption{The orbital-phase dependence of emission and absorption at 0.5 TeV. The blue dashed curve shows the normalized profile of the BBC flux for model 1, which peaks at the superior conjunction due to the anisotropy of Compton scattering. The red dotted curve shows the pair-production attenuation at $z=a$, which is also strongest at the superior conjunction. The black solid curve shows the BBC flux observed from this $z$ (the product of the two functions).
} \label{phi}
\end{figure}

Fig.\ \ref{bc_emissivity}(a) shows the BBC spectra from four values of $\xi$, and Fig.\ \ref{bc_profile} shows the corresponding spatial profiles at 1 GeV and 0.5 TeV. From both figures, we see the BBC emission is dominated by $z\sim 10^3 z_{\rm m}\sim a/2$. This is in agreement with the linear dependence of the BBC flux on the number of scattering electrons in the undiluted blackbody region, i.e., at $z\la a$, and the effect of blackbody dilution at higher $z$. The peak of the 0.5 TeV emission is at a higher $z$ than that for 1 GeV, which is due to the scattering to 0.5 TeV being in the KN regime. Then, the synchrotron cooling is relatively more important at a given $z$ than in the Thomson regime (see Fig.\ \ref{gdotbb}) and a higher $z$ is required for the BBC emission to be strong (while still in the undiluted blackbody density regime).

We then consider the effect of pair absorption on the spectrum. Fig.\ \ref{taugg} shows the pair-absorption optical depth, $\tau_{\gamma\gamma}$ as a function of $E$ for three orbital phases, $\phi_{\rm b}$ (defined to be 0 at the superior conjunction), and $z\simeq a$, around which height the high-energy \g-ray emission peaks, as shown above. We see that $\tau_{\gamma\gamma}$ becomes $>1$ at $E\ga 0.2$ TeV at the superior conjunction. We note that $\tau_{\gamma\gamma}$ decreases rather fast with $z$ (see also \citealt*{rdo10}). At $E=0.5$ TeV and the superior conjunction, $\tau_{\gamma\gamma} \simeq 9.2$, 1.6, 1.0, 0.4 for $z=0$, $a$, $(4/3) a$ and $2 a$, respectively. Also, $\tau_{\gamma\gamma}$ decreases fast with $\phi_{\rm b}$, as shown for $z=a$ by the red dotted curve in Fig.\ \ref{phi}. On the other hand, the unabsorbed BBC component peaks at $\phi_{\rm b}=0$, due to the Compton anisotropy, shown by the blue dashed curve in Fig.\ \ref{phi}. The black solid curve in Fig.\ \ref{phi} shows the absorbed 0.5 TeV flux, which, for the above value of $z$, peaks at an intermediate phase.

\subsection{Hard injection spectrum}
\label{hard}

Subsequently, we assume that the observed MeV tail is due to the jet synchrotron emission (model 2). The resulting spectra are shown in Fig.\ \ref{spectra}(b), and the parameters of the model are given in Table \ref{t:models}. In this model, the injection electron distribution is very flat, and the magnetic field energy density is significantly above the equipartition level. The synchrotron continuum above the turnover energy is due to the cooled electrons, $\gamma_{\rm b0}\ll \gamma_{\rm t0}$. Consequently, the optically-thick part is strongly affected by the height-dependent cooling, resulting in the effective synchrotron spectrum $\alpha>0$ in the radio--to--IR range. 

In this model, very little of the observed emission originates from $z>z_{\rm M}$. The electron power, $P_{\rm e}(z_{\rm M}) \sim 0.1 P_{\rm inj}$, is lost mostly adiabatically at $z>z_{\rm M}$, as in model 1. There is now a substantial contribution to the total BBC power ($\simeq 35$ per cent) from $z>z_{\rm M}$. However, the corresponding contribution to the observed BBC spectrum is small because most of the jet BBC emission is directed back to the star due to the Compton anisotropy. On the other hand, there is a larger contribution from the counterjet, but it is relativistically de-boosted.

The steady-state electron distribution at four values of $\xi$ is shown in Fig.\ \ref{N_gamma}(b). The KN effects are not important here, because of the dominance of the synchrotron cooling. On the other hand, we see a strong effect of the synchrotron self-absorption at low values of $\gamma$ and $\xi$, which reduces the synchrotron cooling and increases $N(\gamma,z)$ to the level corresponding to the adiabatic cooling only. However, this happens at rather low values of $\gamma$, and it does not affect optically-thin synchrotron and Compton spectra. The optically-thick synchrotron spectrum is not affected either as it is approximately independent of $N(\gamma,z)$; see equation (67) in Paper I. At $\gamma\gg \gamma_{\rm t0}$, the distributions are entirely cooled at $\xi\la 10^2$, but become partly uncooled at higher $\xi$. The cooled, high-energy segments of the electron distributions follow $z^2 N(\gamma,z)\propto z$ scaling (proportional to the electron number per unit height; see equations 48--49 of Paper I).

Fig.\ \ref{syn_emissivity}(b) shows the contributions to the total synchrotron spectrum from logarithmic height intervals at increasing $\xi$. We see that the highest $\xi$ dominate at $E\ga 1$ eV. This is a consequence of the assumed form of $Q(\gamma,z)$, which, for $p<2$, yields a slightly increasing integrated injected power per $\ln z$ (Paper I). Fig.\ \ref{syn_z_profile}(b) shows the corresponding vertical profiles of the synchrotron flux at three radio frequencies, dominated by the emission of uncooled electrons. The slopes at $z\leq z_{\rm M}$ follow the same formulae as model 1, see above. However, the emission peaks at $z$ a factor of $\simeq 2$ higher, which requires then a higher wind density than in model 1 in order to account for the observed orbital modulation \citep{zdz12}. 

Fig.\ \ref{bc_emissivity}(b) shows the contributions to the total BBC spectra from logarithmic height intervals at increasing $\xi$. We see that the region at $z\sim 2 a$ ($\xi\sim 10^4$) dominates the BBC emission, except for the TeV region, where even higher values of $z$ dominate. Some offset to higher $z$ with respect to model 1 is a consequence of $z^2 N(\gamma,z) \propto z$ scaling in the model 2, which favours scattering in the highest parts of the jet. 

We also comment here on the full advective solution for $N(\gamma,z)$. We use now the electron distribution of equation (40) of Paper I, giving the analytical advective solution for the case with the dominant synchrotron losses, instead of the local-cooling solution (45). The advective solution does not take into account the influence of self-absorption on $N(\gamma,z)$, which effect is, however, weak (see above). We find the effect of advection on the resulting electron and emission spectra is rather minor. Although $N(\gamma,z=z_{\rm m})\equiv 0$ for solutions with advection, the electron density reaches the cooled form rather quickly. In particular, the height at which the self-absorption optical depth is maximized is $\xi_\tau\simeq 1.028$, i.e., almost equal to $\xi_\tau=1$ of local-cooling models. Table \ref{t:models} lists the parameters of this model, denoted there as 2a. We see these are virtually identical to the corresponding parameters of the local model 2. (The only exception is the value of $\sigma_{\rm eq}$, which formally goes to infinity at $\xi=1$ because $N(\gamma)\equiv 0$ there; the value given in Table \ref{t:models} corresponds to $\xi_\tau$.) This confirms one of the main results of Paper I, namely that the effect of advection is minor for the $Q(\gamma,\xi) \propto\xi^{-3}$ type injection rate in conical jets. 

\begin{figure*}
\centerline{\includegraphics[width=11.5cm]{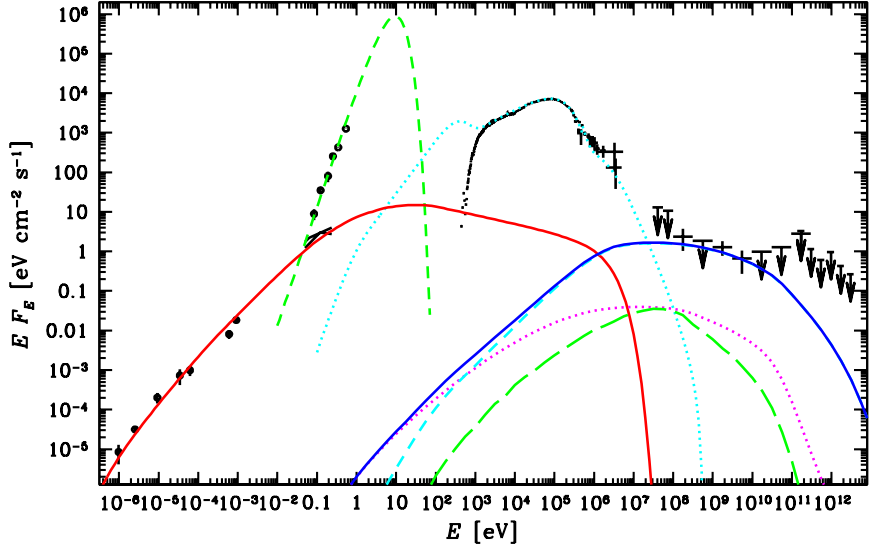}} 
\caption{The hard-state broad-band spectrum of Cyg X-1 for the model 1m. See Fig.\ \ref{spectra} for the description of the data points and the color-coding of model curves.
} \label{spectragmin}
\end{figure*}

\subsection{Low-energy electron cut-off}
\label{gammamin}

Finally, we consider the effect of a high low-energy cut-off in the electron injection function (see Paper I). We now assume the minimum Lorentz factor of the accelerated electrons of $\gamma_{\rm m}=300$. We have calculated the models 1m and 2m, analogous to the previously discussed models 1 and 2 (with $\gamma_{\rm m}=2$), respectively. The corresponding model parameters are given in Table \ref{t:models}. Fig.\ \ref{spectragmin} presents the resulting model 1m spectra (the spectrum of the model 2m is rather similar to that shown in Fig.\ \ref{spectra}b, as can be inferred from Table \ref{t:models}).

Contributions to the synchrotron spectrum from several values of $\xi$ for the model 1m are shown in Fig.\ \ref{syn_emissivity}(c). Similarly to model 2, models 1m and 2m do not show any distinct break at the turnover energy; also the contributions to the observed spectra from $z>z_{\rm M}$ are negligible, similarly to the cases of models 1 and 2.

Overall, the main effect of the increased $\gamma_{\rm m}$ for model 1 is on energetics, with the $P_{\rm i}$ for model 1m lower now by a factor $> 10^2$. Also the values of $\beta_{\rm eq}$ and $\sigma_{\rm eq}$ have significantly changed. For model 2, the changes are minor because the injected electron distribution is hard. 

\subsection{The MAGIC flare}
\label{magic_flare}

\begin{figure}
\centerline{\includegraphics[width=8cm]{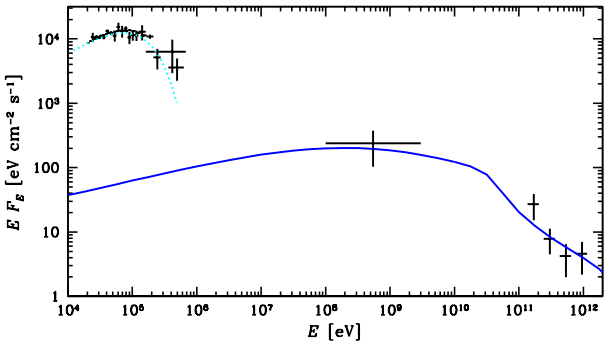}} 
\caption{The simultaneous \integral/MAGIC X-ray--to--\g-ray flare spectrum together with the flare observed by \agile\/ at a different time. The solid curves show the soft injection (pair-absorbed) model spectra associated with an increase of $Q(\gamma)$, as described in Section \ref{magic_flare}. The cyan dotted curve shows the e-folded power-law fit of \citet{malzac08}.
} \label{magic_lowB}
\end{figure}

As we see in Fig.\ \ref{phi}, the emitted BBC flux is strongest at the superior conjunction, due to the Compton anisotropy. This may explain the orbital phase of $\phi_{\rm b}/2\upi=0.90$--0.91 of the flare observed by the MAGIC telescope in the 0.2--1 TeV energy range on MJD 54002.928--982 \citep{magic}, provided the pair opacity is low. The MAGIC spectrum is shown in Fig.\ \ref{magic_lowB} together with the simultaneous hard X-ray spectrum from \integral, which revealed an increase by a factor of $\sim 3$ with respect to the average hard-state spectrum \citep{malzac08}. Using the formalism of Paper I, the optical depth for the 0.5 TeV photons is $\tau_{\gamma\gamma}<1$ at $z\ga 1.3 a$. Thus, the MAGIC flare can possibly be explained if most of the TeV emission took place at such heights, as earlier found by \citet{rdo10}.

Unfortunately, the available broad-band data at the time of the MAGIC flare are rather sparse, and do not allow for a unique determination of the model parameters. In particular, there was neither radio nor GeV monitoring around this event. Observations simultaneous with MAGIC are available only in the X-ray band, see the \swift-BAT and \xte-ASM data in \citet{magic}, and \integral\/ data in \citet{malzac08}. However, since production of X-rays is strongly dominated by the accretion flow, those data do not provide spectral constraints on the models of the TeV flare production in a jet. Still, the found coincidence of the TeV flare with the observed increase of the X-ray activity strongly suggests that the TeV flare resulted from a temporary change of the conditions at the jet base. Since the duration of the flare, $\sim 5000$s, is longer than the observer-frame time of propagation of the jet over the region where most of BBC radiation is produced, $z\sim a$, such an event can be approximated by a steady-state model. 

We thus consider two simple scenarios, representing modifications of the soft and hard injection models. In the first one, we temporarily increase the electron injection rate. In the second scenario, we assume that some unspecified process decreases temporarily the jet magnetic energy flux, and increases the amplitude of the electron distribution; this results in a temporarily increased rate of the Compton scattering. In both cases, we keep all other parameters fixed as before. Naturally, more than one parameter could have changed in reality, including, e.g., $\eta_{\rm acc}$ in addition to $Q(\gamma)$ and $B$. The considered scenarios are therefore only the two examples used to investigate if the developed jet model can account for the TeV flare of Cyg X-1.

In our soft-injection model 1, which we now consider as the underlying steady state during which the TeV flare happened, the magnetic field is rather weak. Then, the energy losses at high $\gamma$ and $z$ (which are responsible for the TeV emission) are almost completely dominated by the adiabatic and BBC processes. Thus, there will be a unique shape of the high-energy BBC spectrum, dependent only on the form of the acceleration function, $Q(\gamma)$. The \g-ray flux of that spectrum is then $\propto Q(\gamma)$, and independent of the model magnetic field. Therefore, this scenario requires a temporary increase of the acceleration rate in order to explain the flare. This could have happened, e.g., due to an interaction of the jet with a clump in the stellar wind of the companion \citep{rdo10}. Below we consider a simple model assuming an increase of the injection normalization, $Q_0$. We note, however, that the increase could involve only the high-energy segment of the electron distribution (at $\gamma\ga 10^5$), which is responsible for the production of the $\geq 0.2$ TeV photons. 

In the local approximation, the steady-state electron distribution is thus given by equation (\ref{N_cool}) with $\dot\gamma\simeq \dot{\gamma}_{\rm ad}+\dot\gamma_{\rm BBC}$. Then, equations (70), (74) of Paper I give us the emitted BBC spectrum at the orbital phase of the flare, which then is subject to pair absorption. The resulting spectrum at $Q_0$ increased by a factor of 100 at the orbital phase of $\phi_{\rm b}/2\upi=0.9$ is shown in Fig.\ \ref{magic_lowB}. We see it reproduces well the MAGIC data. We also show the flux of the hard-state flare reported at a different time by \agile\/ \citep{sabatini10}. We see that our model reproduces that data as well. A caveat for this scenario is the required high amplitude of the increase of the acceleration rate.

\begin{figure*}
\centerline{\includegraphics[width=11.5cm]{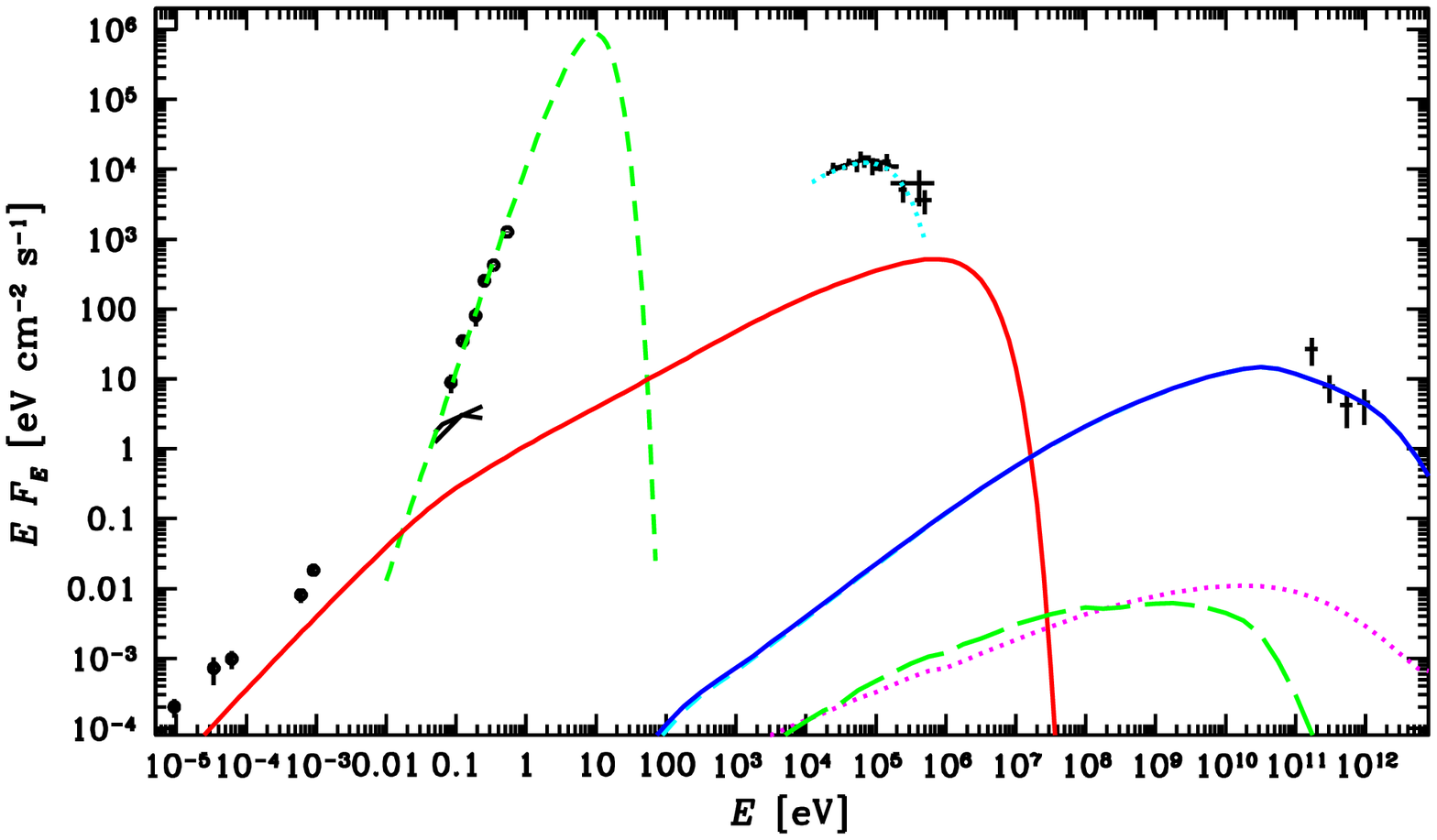}} 
\caption{The simultaneous \integral/MAGIC X-ray--to--\g-ray spectrum of Cyg X-1 (\citealt{malzac08,magic}; shown by crosses at $\geq 10$ keV) together with the average hard-state radio/IR spectrum. The solid curves illustrate the hard-injection (pair-absorbed) model for the MAGIC flare (associated with a decrease of $B$) described in Section \ref{magic_flare}. The meaning of other symbols is the same as Fig.\ \ref{spectra} except for the cyan dotted curve showing now the e-folded power-law fit of \citet{malzac08} to the \integral\/ spectrum.
} \label{fig:magic}
\end{figure*}

On the other hand, the increase of the $\gamma$-ray flux due to a reduction of $B$ can happen in models in which cooling of electrons with $\gamma\sim 10^5$ is dominated by the synchrotron process. This is the case for our models 2 and 2m. Fig.\ \ref{fig:magic} shows a modified model 2 (at $\phi_{\rm b}/2\upi=0.9$) reproducing the MAGIC flare spectrum. Here the magnetic field is reduced by a factor of several, down to $B_0=9\times 10^4$ G, whereas all the other parameters remain unchanged. There is a slight increase of the total power due to the increase of $\gamma_{\rm M}$ related to the reduction of $B$ (see equation 6 in Paper 1). The radio/IR flux has decreased by a factor of several. The flux at $>0.2$ TeV is almost completely due to the emission produced at $z\ga a$. Therefore, the effect of pair absorption is moderate, reducing the flux at $>0.2$ TeV by at most $\sim 30$ per cent.

We note that the Alfv{\'e}n time scale, $t_{\rm A} \equiv z \tan\Theta_{\rm j}/v_{\rm A}$, where $v_{\rm A}=c (1+1/\sigma_{\rm eq})^{-1/2}$ is the Alfv{\'e}n speed, can be rather short. In model 2, $v_{\rm A} \simeq c$ and hence $t_{\rm A} \simeq 5$ s at $z=a$ (in the observer's frame). Also, the cooling time of $\gamma\ga 10^5$ electrons (responsible for upscattering of stellar blackbody photons to TeV energies) is $\la 15$ s in this model at $z=a$. Both times scales are therefore shorter than the dynamic time at this height, which is $\simeq 180$ s, and significantly shorter than the duration of the MAGIC flare of $\simeq 5000$ s. The dynamical time scale is instead compatible (in both models) with the observed TeV flux rise after the first half of the observation \citep[during which no TeV signal was detected;][]{magic}.

An observational test between the two scenarios may be provided by radio measurements simultaneous with a possible future TeV flare. In the first scenario, the electron injection rate strongly increases, which would increase the radio fluxes (unless the increase occurs only at the highest values of $\gamma$, in which case the radio flux would remain constant). In contrast, a decrease of $B$ in the second scenario leads to a significant decrease of the radio fluxes, as shown in Fig.\ \ref{fig:magic}. On the other hand, the reconnection event postulated in our model 2 might have led to an increased electron acceleration rate. Thus, the actual flare could have happened due to a combination of our two scenarios. 

\section{Discussion}
\label{discussion}

In the framework of our model, the electron energy losses are compensated by a continuous injection of freshly accelerated particles along the outflow, with the assumed constant rate per unit $\gamma$ and unit $\ln z$. Then, there is an issue of a source of the energy injected in the jet through the particle acceleration process. One possibility is the energy associated with colliding shells at different bulk Lorentz factors, $\Gamma_i$ (where the index $i$ numbers the shells) in the internal shock model \citep*{jfk10,malzac13}. In this case, electrons are likely accelerated via the diffuse 1st-order Fermi process operating at shock fronts. Another possibility is the magnetic energy source, in which case electrons may be accelerated directly via magnetic reconnection and reconnection-driven turbulence. 

In the first scenario, our assumed jet bulk Lorentz factor, $\Gamma_{\rm j}$, should be interpreted as the average of the shell Lorentz factors, $\langle\Gamma_i \rangle$. The available energy is, on the other hand, related to the Lorentz factor dispersion, $\Delta\Gamma_i$. Assuming $\Delta\Gamma_i\sim \langle\Gamma_i \rangle$ (which is likely but not strictly required), the available power is comparable to that of the jet kinetic luminosity $P_{\rm i}$, which requires $P_{\rm i}\ga P_{\rm inj}$ in order to account for the observed emission and associated adiabatic losses. Table \ref{t:models} gives the values of $P_{\rm i}$ corresponding to cold ions associated with relativistic electrons. Since ions may be not cold in the jet rest frame, and since there are likely to be more ions than that implied by the number density of relativistic electrons, the values provided represent lower limits only. In the model 1, the above condition is always satisfied, and thus electrons can be readily accelerated in shocks. In the other models, the given lower limits on the ion power are $<P_{\rm inj}$, and thus a large number of additional ions is required for the shock model to be considered as plausible. Such higher values of $P_{\rm i}$ would also be consistent with the estimates of the jet kinetic power of \citet{gallo05} and \citet{russell07}. 

On the other hand, for the models 2 and 2m we obtain $P_B>P_{\rm inj}$, and thus the reconnection-driven electron acceleration is plausible in those models. We also note that all of the components of the jet power for our models (see Table \ref{t:models}) are significantly below the luminosity of the accretion flow in the hard state. Thus, the jet can be powered by accretion in all the considered cases. 

A related diagnostic is the value of the magnetization parameter, $\sigma_{\rm eq}$. In order to enable a formation of strong hydrodynamic shocks dissipating efficiently bulk kinetic energy of the outflow to nonthermal particles, low or at most only moderate plasma magnetization is allowed, $\sigma_{\rm eq} \la 1$ \citep{ss09,lyubarsky10b,mimica10}; this condition is satisfied in models 1 and 1m. On the other hand, $\sigma_{\rm eq} \ga 1$, as in our models 2 and 2m, is required for the efficient magnetic-to-particle energy transfer via magnetic reconnection process (\citealt{lk01}; \citealt{lyubarsky05}; \citealt{lyubarsky10a} and references therein; \citealt*{kms13}; \citealt{ss14}).

Table \ref{t:models} also gives the total (integrated over the jet volume) rate of injection of relativistic electrons, $R_{\rm inj}$, and the lower limit on the flux of the electrons through the jet, $R_{\rm e}$ (see Paper I). This lower limit corresponds to the relativistic electrons only, whereas there can be `cold' electrons as well. If $R_{\rm inj}$ exceeds the actual value of $R_{\rm e}$, electrons need to be accelerated more than once. Although $R_{\rm inj} > R_{\rm e}$ is the case for all the models considered, the two numbers are close to each other. Thus, only a moderate number of `cold' electrons suffice for acceleration being a single event for an individual electron.

We stress that our results depend on the value of the jet opening angle, $\Theta_{\rm j}$, which we have assumed at the upper limit of \citet{stirling01} of $=2\degr$. If it is lower, it will change the obtained $z_{\rm m}$ and $B_0$, approximately satisfying $z_{\rm m}/B_0^{1/4}\propto \Theta_{\rm j}^{-1/2}$ (e.g., eq.\ 22 in \citealt{zls12}). However, $B_0$ is the major factor determining the flux in the BBC component, and thus it is not constant with changing $\Theta_{\rm j}$. For example, assuming $\Theta_{\rm j}=0.5\degr$ for the model 1 and requiring the \g-ray flux to reproduce the \fermi\/ data result in $\xi_{\rm m}\simeq 1400$ and $B_0\simeq 6\times 10^3$ G. Also, $\beta_{\rm eq}$ increases by a factor of $\sim 10$ and $\sigma_{\rm eq}$ and $P_B$ decreases by the same factor. The other parameters change only slightly. 

The average broad-band spectrum of Cyg X-1 in the hard state studied by us (see Fig.\ \ref{spectra}) includes the IR measurement of the synchrotron turnover, $E_{\rm t0}$, from \citet{rahoui11}. We can see that the total IR flux (dominated by the stellar blackbody and the wind emission) in the energy range fitted by \citet{rahoui11} is much larger than their claimed jet component. Also, \citet{rahoui11} used a broken power law in their fits whereas theoretical considerations point to a rather gradual curvature (e.g., \citealt{hj88,zls12}), as supported by a number of observational findings \citep{russell13}. Indeed, \citet{russell13} point out the uncertainty of that spectral break measurement. Furthermore, there is a possible IR contribution from synchrotron emission of hybrid (thermal+non-thermal) plasma in the accretion flow \citep{vpv11,vpv13}, which might affect the non-stellar fluxes measured in that region.  Our model parameters do depend relatively sensitively on the actual value of $E_{\rm t0}$; e.g., we obtain $\xi_{\rm m}\simeq 360$, $B_0\simeq 1.7\times 10^4$ G in the model 1 with $E_{\rm t0}=0.3$ eV. However, our general conclusions do not depend on it, as long as $E_{\rm t0}$ is located around IR/optical frequencies. 

Another uncertainty is the origin of the MeV tail, which can be either from the jet or the accretion flow. As discussed in Section \ref{intro}, at the moment there is no compelling statistical evidence for the former possibility. Therefore, we have discussed the jet models which may account for the observed MeV tail (models 2 and 2m), along with the models which are characterized by negligible MeV fluxes (models 1 and 1m).

We also point out the fact that the observed synchrotron emission below the turnover frequency is contributed by both optically-thick and optically-thin regions (e.g., \citealt{hj88}). Given the slopes of this emission with the height (see Fig.\ \ref{syn_z_profile}), most of that emission is from optically-thin regions. This invalidates the polarization model of Cyg X-1 of \citet{rs14}, who assumed that the radio emission of Cyg X-1 originates entirely in synchrotron self-absorbed medium. Since the maximum polarization degree in the optically-thick regime is much lower than that of optically-thin synchrotron emission \citep{ps67,sw68}, \citet{rs14} concluded that this is a possible reason for the low linear polarization of the source at 5 GHz \citep[with the derived upper limit $< \simeq 8$ per cent;][]{stirling01}. This, in turn, allowed their model polarization of the fully optically-thin synchrotron emission to be high (almost 100 per cent, see fig.\ 1 therein). However, since the 5 GHz emission originates mostly from optically-thin parts of the jet, as shown in Fig.\ \ref{syn_z_profile}, the low upper limit on its polarization implies that the optically-thin synchrotron emission is also weakly polarized, with a similar upper limit on the polarization as that measured at 5 GHz. 

These considerations allow, in fact, to rule out our models 2 and 2m as a viable explanation of the very high MeV polarization in Cyg X-1 claimed by \citet{l11} and \citet{jourdain12}. In those models, the bulk of the observed MeV emission originates from the $\xi\ga 10^3$ regions (see Fig.\ \ref{bc_emissivity}), where the 5 GHz emission is optically thin, and at most weakly polarized. If the MeV polarization is therefore real, more complex jet models are required. In this context, the operation of the Soft Gamma-ray Detector on board the {\it ASTRO-H\/} satellite \citep{tak12}, enabling hard X-ray polarimetry of bright celestial sources, is much anticipated.

We also comment here on the model of \citet*{romero14}, in which the MeV polarization originates in a nonthermal corona around the black hole. It appears that their equation (2) is in error. It gives the fractional polarization from a magnetic field configuration with an axial symmetry, following the results of \citet{ks62}. The fractional polarization is given by \citet{romero14} as a function of the field components independent of the viewing angle of the system. However, this appears impossible in the case of axial symmetry, in which, e.g.,  viewing the system along the axis would result in null net polarization. Indeed, the corresponding equations (35--36) in \citet{ks62} are given in terms of the field components {\it projected on the sky\/} rather than the intrinsic ones. Taking this into account introduces an additional factor of $\sin^2 i$ in equation (1) of \citet{romero14}, which reduces the polarization by a factor of 4.3 for our assumed inclination, which is clearly much below the fractional polarization claimed by \citet{jourdain12}. This appears to rule out that model.

Our vertical profiles of radio emission, Fig.\ \ref{syn_z_profile}, can be tested against jet radio maps of Cyg X-1. \citet{stirling01} found that $\sim 1/3$--1/2 of the emission at 8.4 GHz is unresolved, with the resolution of 3 mas, corresponding to $\simeq 50 a$ for the distance and jet inclination adopted here. Thus, our model 1 is characterized by a deficit radio emission above $50 a$, what requires an additional source of the energy dissipation at large radii. On the other hand, model 2 appears approximately consistent with the radio observations. 

We have found that the BBC component (i.e., scattering of stellar blackbody photons) strongly dominates over the SSC and XC ones in high-energy \g-rays. This predicts a relative weakness of high-energy \g-ray emission from low-mass X-ray binaries. So far, this appears to be confirmed by the lack of a detection of any such binary in the GeV range.

\section{Conclusions}
\label{conclusions}

In this paper, we have applied the jet model developed in Paper I to the average broad-band spectrum of Cyg X-1 in the hard state. We take into account the new measurements and upper limits in the 30 MeV--300 GeV range by \fermi-LAT of MZC13. We have found that in $\gamma$-rays, the Compton scattering of blackbody photons dominates over the SSC process, and that the resulting evaluated jet emission accounts well for the observed LAT spectrum of the source.

In the context of the recent controversial claim of a strong polarization of Cyg X-1 at MeV photon energies, we have considered two different variants of the developed jet model. In one, the observed MeV tail is assumed to originate in the accretion flow. This model returns `standard' parameters including soft injection index of the radiating electrons, $p \ga 2$, and the jet magnetic field which can be close to the equipartition level. In the other model, the MeV tail is accounted for by the jet synchrotron emission, so that its strong (maximum) polarization is not, in principle, implausible. We find that in this case the \fermi-LAT measurements impose very tight constraints on the jet magnetic field, which has to be significantly above the equipartition level in order to avoid an overproduction of GeV-energy $\gamma$-rays. Also, the electron injection index has to be hard, $p\sim 1.5$. Importantly, however, the bulk of the MeV jet synchrotron emission turns out to originate in the same regions as the jet radio emission. The strong observational upper limit on the radio linear polarization ($\la 8$ per cent) rules therefore out any strong MeV polarization in our model.

We have also modelled the TeV flare of Cyg X-1 observed by MAGIC \citep{magic}. We have speculated that it could have been due to a magnetic reconnection event temporarily decreasing the jet magnetic field, but only assuming hard electron injection. Alternatively, it could have been due to a temporary increase of the acceleration rate by a factor of, at least, $\sim 10^2$, around $\gamma\ga 10^5$, in models with soft electron injection.

Finally, in Appendix \ref{advection}, we have derived equations describing the electron distribution in jet regions above the dissipation region, where the distribution is governed solely by the advection and radiative losses. These equations allow for a very simple treatment of jet models with acceleration taking place only close to the jet base \citep{kaiser06,pc09}, in which case advection is of a crucial effect at higher heights. This provides an important extension to the results of Paper I.

\section*{ACKNOWLEDGMENTS}

We are grateful to the referee for careful reading of the manuscript and valuable comments. We also thank G. Romero and D. Russell for stimulating comments. This research has been supported in part by the Polish NCN grants 2012/04/M/ST9/00780, DEC-2012/04/A/ST9/00083 ({\L}.S.) and DEC-2011/01/B/ST9/04845 (M.S.).

\appendix

\section{Electron transport beyond the dissipation region}
\label{advection}

The evolution of $N(\gamma,z)$ at $z>z_{\rm M}$ is given by equation (\ref{ndot}) (or, more generally, equation 24 in Paper I) with null injection, $Q(\gamma,z)\equiv 0$, and with the boundary condition at $z_{\rm M}$ given by $N(\gamma, z_{\rm M})$ obtained within the dissipation region. The general solution to this equation in the notation of Paper I is found to be (cf.\ \citealt{stawarz08})
\begin{eqnarray}
\lefteqn{
\tilde{N}(\gamma, \xi) = \label{general}}\\
\lefteqn{\cases{\displaystyle{
\exp\left\{\!-\!\!\int^\xi_{\xi_{\rm M}}\!\!\!\! {\rm d}\xi' {\partial\gamma_\xi[\gamma'(\gamma,\xi;\xi'),\xi']\over \partial \gamma'} \right\}\! \tilde{N}\left[\gamma'(\gamma,\xi; \xi_{\rm M}), \xi_{\rm M}\right]\!,}& $\gamma'<\gamma_{\rm max}$;\cr
0 & $\gamma'\geq\gamma_{\rm max}$,}
\nonumber}
\end{eqnarray}
where $\gamma_{\rm max}$ is the maximum Lorentz factor in the distribution $\tilde{N}(\gamma',\xi_{\rm M})$, and $\gamma'$ ($>\gamma$), the Lorentz factor at $\xi_{\rm M}$ corresponding to $\gamma$ at $\xi$, is given by the solution of equation (28) in Paper I. The physical reason for the above condition is that the cooling can be so fast that there is no Lorentz factor $\gamma'$ (even at $\gamma_{\rm max}\rightarrow \infty$) at $\xi_{\rm M}$ that can be cooled to a $\gamma$ at $\xi$, see, e.g., fig.\ 3 in Paper I. Equation (\ref{general}) is the counterpart of equation (29) in Paper I for $\xi>\xi_{\rm M}$.

As noted in Paper I, much simpler solutions of the transport equation can be obtained if the radiative losses are synchrotron and Thomson. For a general dependence of the radiation + magnetic energy density on $\xi$, $\gamma'(\gamma, \xi; \xi_{\rm M})$ is given by equation (32) in Paper I. The solution for the electron distribution in the notation of Paper I for $\xi>\xi_{\rm M}$ (and $\gamma\gg 1$) can be obtained from equation (\ref{general}) using equations (A5--A7) of \citet{stawarz08}. It is
\begin{equation}
\tilde{N}(\gamma,\xi)=\cases{\displaystyle{{\gamma'^2 r_{\rm j}(\xi_{\rm M})^{2/3}\over \gamma^2 r_{\rm j}(\xi)^{2/3}} \tilde{N}(\gamma',\xi_{\rm M}),} & $\gamma'<\gamma_{\rm max}$;\cr
0 & $\gamma'\geq\gamma_{\rm max}$,}
\label{A8}
\end{equation}
where $r_{\rm j}(z)$ is the local jet radius. The above equation is the counterpart of equation (33) in Paper I for $\xi>\xi_{\rm M}$.

If the jet is conical and has a constant speed, equation (\ref{A8}) becomes
\begin{equation}
N(\gamma,z)=\cases{\displaystyle{\left(\gamma'\over \gamma\right)^2 \left(z\over z_{\rm M}\right)^{-8/3} N(\gamma',z_{\rm M}),} & $\gamma<\gamma_{\rm h}(z)$;\cr
0 & $\gamma\geq\gamma_{\rm h}(z)$,}
\label{N_adv}
\end{equation}
where $\gamma_{\rm h}(z)$ is the highest possible $\gamma$ in $N(\gamma,z>z_{\rm M})$ corresponding to $\gamma'<\infty$. Since $z_{\rm M}\gg a$, the dependence of the stellar radiation energy density on $z$ is $\propto z^{-2}$, the same as that of the magnetic one for $B\propto z^{-1}$. In that case, we obtain an analytic solution for $\gamma'$ and $\gamma_{\rm h}$,
\begin{eqnarray}
\lefteqn{
\gamma'(\gamma,z; z_{\rm M})={\gamma (z/z_{\rm M})^{2/3}\over 1-{2\gamma z_{\rm m}\over 5\gamma_{\rm b0} z} \left[(z/z_{\rm M})^{5/3}-1\right]},
\label{gamma_xi}}\\
\lefteqn{
\gamma_{\rm b0}\equiv {A_{\rm ad}\over A_{\rm S}+(a/z_{\rm m})^2 A_{\rm BBC}},
\label{gamma_b}}\\
\lefteqn{
\gamma_{\rm h}(z)=
{5\gamma_{\rm b0} z\over 2 z_{\rm m}}\left[\left(z\over z_{\rm M}\right)^{5/3} -1\right]^{-1},
\label{gamma_max}}
\end{eqnarray}
where $A_{\rm ad}$, $A_{\rm S}$, $A_{\rm BBC}$ are defined by equations (9), (12), (14) in Paper I, respectively, $A_{\rm BBC}$ is to be calculated with the Doppler factor at $z\gg a$ and the unit KN correction, $\gamma_{\rm b0}(z/z_{\rm m})$ is the Lorentz factor of the radiative cooling break at $z$. The jet terminates at $z_{\rm h}$ for which $\gamma_{\rm h}(z_{\rm h})=\gamma_0$. Equations (\ref{gamma_xi}--\ref{gamma_max}) represent the counterparts of equations (35--37) in Paper I for $\xi>\xi_{\rm M}$. With the obtained distribution, we can calculate the spectral contributions and the power lost adiabatically at $z>z_{\rm M}$. 

\label{lastpage}

\end{document}